\pdfoutput=1 
\documentclass{JINST}
\usepackage{amsmath}
\usepackage[]{units}

\title{Prototype muon detectors for the AMIGA component of the Pierre Auger Observatory}

\author{
\par\noindent
\bf The Pierre Auger Collaboration\\
The Pierre Auger Observatory,\\
Malarg\"ue, Argentina

}

\abstract{AMIGA (Auger Muons and Infill for the Ground Array) is an upgrade of the Pierre Auger Observatory to extend its range of detection and to directly measure the muon content of the particle showers. It consists of an infill of surface water-Cherenkov detectors accompanied by buried scintillator detectors used for muon counting. The main objectives of the AMIGA engineering array, referred to as the Unitary Cell, are to identify and resolve all engineering issues as well as to understand the muon-number counting uncertainties related to the design of the detector. The mechanical design, fabrication and deployment processes of the muon counters of the Unitary Cell are described in this document. These muon counters modules comprise sealed PVC casings containing plastic scintillation bars, wavelength-shifter optical fibers, 64 pixel photomultiplier tubes, and acquisition electronics. The modules are buried approximately \unit[2.25]{m} below ground level in order to minimize contamination from electromagnetic shower particles. The mechanical setup, which allows access to the electronics for maintenance, is also described in addition to tests of the modules' response and integrity. The completed Unitary Cell has measured a number of air showers of which a first analysis of a sample event is included here.}


\dedicated{\textnormal{FERMILAB-PUB-16-164-AD-AE-CD-TD}}

\begin{document}

\section{Introduction}

The Pierre Auger Observatory \cite{pao_nim}, located in the province of Mendoza, Argentina, is a hybrid detector covering \unit[3,000]{km$^2$} with 1660 surface stations (the surface detector, SD) and 27 fluorescence telescopes. The SD stations are arranged in a triangular grid of mostly \unit[1.5]{km} spacing, while the telescopes are split amongst four sites at the edge of the surface array. Currently, the Auger Observatory is being upgraded, and AMIGA \cite{amiga_icrc} is one of the enhancement projects. 

Two of the main objectives of AMIGA are to measure the composition-sensitive observables of extensive air showers and to study features of the hadronic interactions. The optimum set of detector measurements for an event includes energy, atmospheric depth at shower maximum X$ _{max,EM}$, and the number of muons at an optimal lateral distance from the shower axis N$_{\mu}^{opt}$. Additionally, the muon production depth X$ _{max,\mu}$ (the reconstruction of which is under development) is an observable of considerable interest. Reconstruction of these parameters requires quality measurements of the muonic and the electromagnetic components of showers. Once these parameters are measured, the best available multi-parametric fit may be applied. AMIGA is designed to measure muon arrival times and N$_{\mu}^{opt}$. The former might enable the reconstruction of the muon longitudinal profile. The latter is obtained by sampling the muon lateral distributions. These two observables reduce energy reconstruction systematics, particularly when the fluorescence detector is not operating. Note that AMIGA can measure the muon longitudinal profile with a nearly 100\% duty cycle.

\begin{figure}[tp]
   \centering
   \includegraphics[width=.56\textwidth]{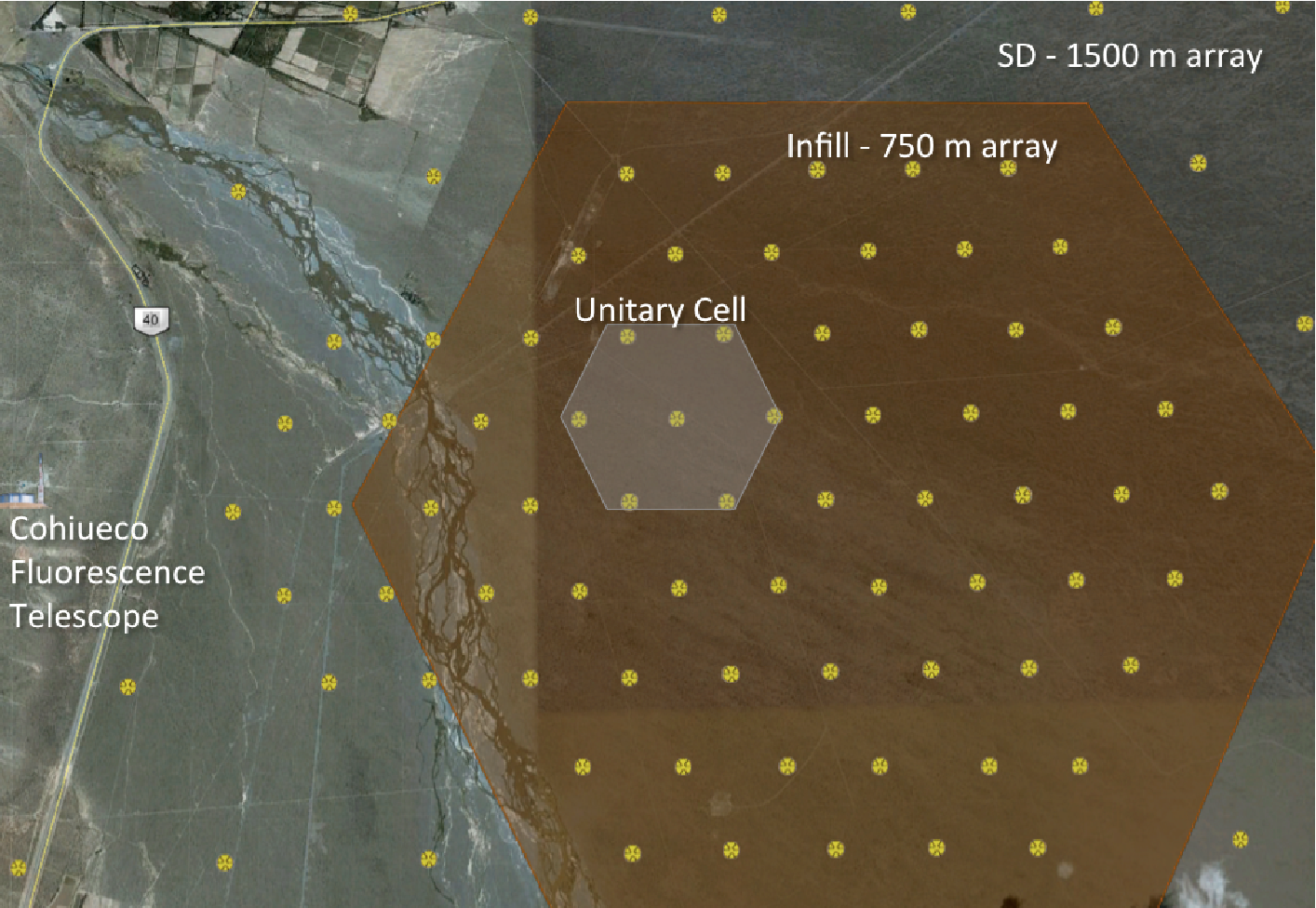}   
   \includegraphics[width=.43\textwidth]{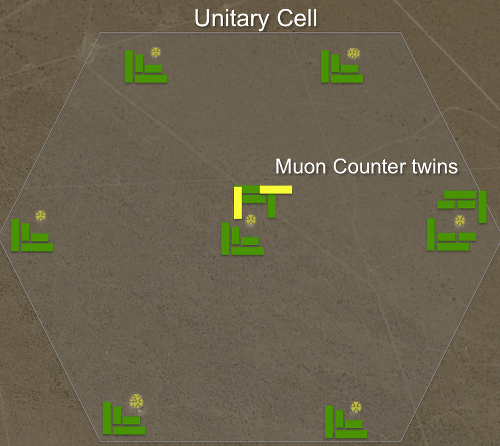}   
   \caption{Left: Map of the AMIGA array with brown background (from \cite{icrc:2013}) with the Unitary Cell engineering array position (in gray). Right: layout of the Unitary Cell showing the locations of MCs and corresponding water-Cherenkov stations. Figure is not to scale. Green and yellow boxes represent MC modules buried at approximately \unit[2.25]{m} and \unit[1.3]{m}, respectively.}
   \label{mapaAugerAMIGA}
\end{figure}

AMIGA consists of an infilled area of 61 detector pairs (fig. \ref{mapaAugerAMIGA}) each one composed of a surface water-Cherenkov detector and a buried \unit[30]{m$^2$} Muon Counter (MC). The AMIGA MCs are deployed on a \unit[750]{m} triangular grid to directly measure the muon content of showers with primary energies greater than \unit[$3\,\times\,10^{17}$]{eV}. A spacing of \unit[750]{m} and an area of \unit[23.5]{km$^2$} were chosen due to the small particle footprint and high flux of low-energy showers.

A Unitary Cell (UC) of seven SD stations on a hexagonal grid has already been equiped with MCs in the same area as the Auger Engineering Radio Array (another enhancement of the Auger Observatory) allowing for combined analyses \cite{aera}. The UC has aided in the debugging of engineering issues and the understanding of the counting uncertainties, in order to attain a stable detector and final design for the production of AMIGA. Additionally, one of the most important tasks of the given prototype of muon detectors is to minimize muon-number counting uncertainties stemming from mechanical design. 

\begin{figure}[bp]
\begin{centering}
\includegraphics[width=.9\textwidth]{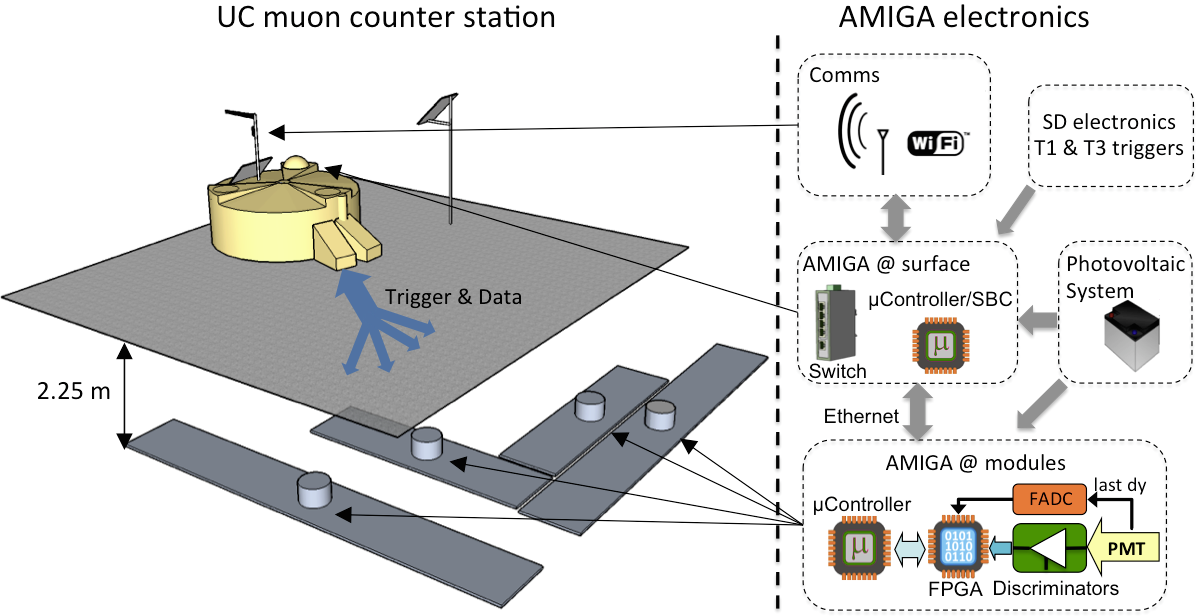}
\par\end{centering}
\caption{Schematics of a Muon Counter (MC) and Surface Detector (SD) of the AMIGA Unitary Cell together with the AMIGA electronics.}
\label{fig:estacion}
\end{figure}

The MCs have a modular design in which the \unit[30]{m$^2$} detection area is divided in two modules with \unit[5]{m$^2$} and two with \unit[10]{m$^2$} detection area. Each of the detector modules has its own acquisition system triggered by the SD station  (fig. \ref{fig:estacion}). Currently, two of the UC position have twin muon detectors, which consist of two \unit[30]{m$^2$} MCs separated by approximately \unit[10]{m}. The purpose of these detectors is to study counting fluctuations. 

\section{Requirements for the AMIGA muon detectors}

Among the many requirements for the MCs, only those affecting the mechanical design of the detectors are mentioned here: 

\begin{itemize}
\item \textit{Number of detectors:} AMIGA is designed to have at least 61 MCs. Therefore, MCs must be easy and fast to manufacture in order to complete the project in a reasonable time period. 
\item \textit{Muon sensitivity:} The MCs must be sensitive only to the muon content of the particle showers produced by incident cosmic rays. Simulations show that a vertical shielding of \unit[540]{g cm$^{-2}$} clearly suffices to reduce the electromagnetic punch-through to a negligible level at core distances of interest. This shielding is equivalent to burying the detectors \unit[$\sim2.25$]{m} underground (considering an average local soil density of \unit[2.38 $\pm$ 0.05]{g cm$^{-2}$} \cite{estudio_suelos}). 
\item \textit{Detection area and segmentation:} Based on simulations, the detection area was chosen to be \unit[30]{m$^2$} divided in 192 segments (see \cite{paper_daniel} for details about detection area and segmentation simulations). The segmentation of the detector is also related to the time span over which a muon is to be sought, since two muons arriving in the same time window could be counted as one. The finer segmentation corresponds to less muon pile-up in a fixed time window. The minimum muon time window is defined by the convolution of the decay time of the scintillator and the optical fiber. 
\item \textit{Lifetime:} Because only the electronics can undergo maintenance over a period of 10 years underground, the scintillator module components' ruggedness and aging characteristics have to be considered. 
\item \textit{Transportation and deployment:} the modules must be both easy to handle and durable enough to withstand both long transportation in a semi-truck trailer as well as hard underground deployment conditions (high pressure from the soil above the modules). They must also have light-tight and water-tight seals. 
\item \textit{Power supply:} Detectors are deployed in the field over a large-area array so a photovoltaic system is the only reasonable power solution. Accordingly, a low-power electronics design must be developed for the production phase of the project.
\item \textit{Costs:} The cost must be as low as possible given the number of scintillator modules to be built. The mechanical design of the modules must be optimized to allow for the use of standard trucks and cranes. 

\end{itemize}

\section{Description of the scintillator modules}
Four modules, \unit[$2\,\times\,5$]{m$^{2}$} and \unit[$2\,\times\,10$]{m$^{2}$}, were deployed at each site in the UC \cite{icrc2011:modulos} for pile-up and signal attenuation cross comparison tests. Additionally, the electronics of the modules allow for both integration of the total detected signal as well as independent counting of muons. Integration allows for measurements close to the core, and the segmentation (scintillation bars) results in a sturdy muon counting technique with low muon pile-up in a defined inhibition time window. 

\begin{figure}[tp]
\begin{centering}
\includegraphics[width=.65\textwidth]{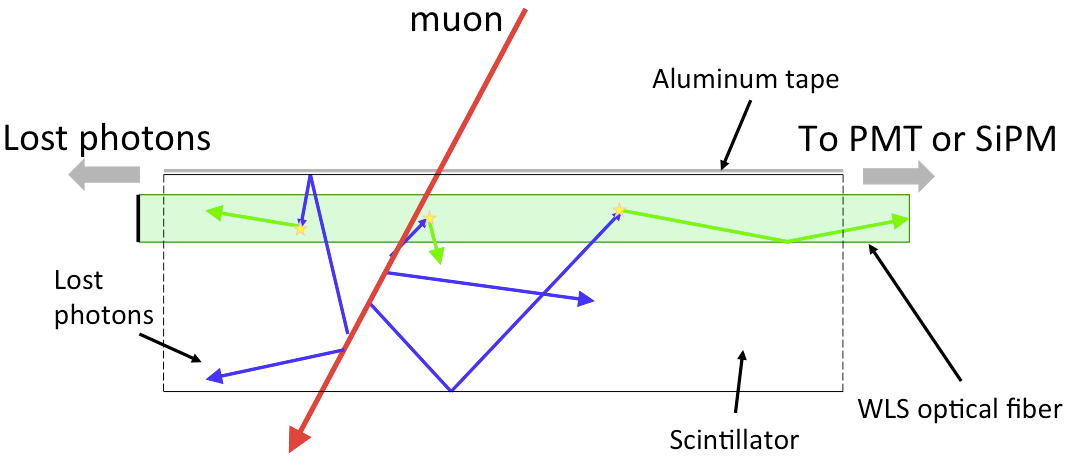}
\par\end{centering}          
\caption{Schematic of a scintillator bar excited by a muon. Highlighted are the trajectories of the incident particle (in red), the photons produced within the bar (in blue) and within the fiber (in green).}
\label{centellador}
\end{figure}

Every scintillator module comprises 64 scintillation bars, each of dimensions \unit[40]{mm}$\,\times\,$\unit[10]{mm} $\,\times\,$\unit[4]{m}, with a \unit[1.2]{mm} diameter wavelength-shifting (WLS) optical fiber glued into a lengthwise groove of the bar. The light produced in the scintillation bars is collected and propagated along the WLS fibers (fig. \ref{centellador}), which then couple to multi-pixel PMTs (photomultiplier tubes). The 64 scintillators and optical fibers are lodged within a PVC (Polyvinyl Chloride) casing and, together with the electronics kit, form the detector module.

The electronics of the modules facilitate identification of pulses above a given threshold and thus allow for the counting of muons without detailed knowledge of signal structure and peak intensity. The thresholds of the module's 64 channels can be individually monitored and set manually or through a calibration algorithm (see \cite{icrc2013:ale} for details about the electronics slow-control implementation). This method is very robust since it does not rely on deconvoluting the number of muons from an integrated signal. Thus, it does minimally depends on the PMT gain (and its fluctuations), the muon impact position on the scintillator bar, and the corresponding light attenuation along the fiber length. It also does not require thick scintillators to reduce Poisson fluctuations in the number of SPEs (single photoelectrons) produced per incident muon. However, it does rely on fine counter segmentation to prevent under counting due to simultaneous muon arrivals. It additionally depends on the adjustment of thresholds within an appropriate range to ensure good counting efficiency.

\subsection{Scintillation bars}
Plastic scintillators were chosen because of their quality mechanical and maintenance-free properties. The scintillation bars, which were produced and quality-controlled at Fermilab, are \unit[4]{m} long extruded bars (for the \unit[10]{m$^2$} modules) of polystyrene doped with fluor and co-extruded with TiO$_2$ as an outer layer for reflectivity. They have a detection cross section of \unit[40]{mm}$\,\times\,$\unit[10]{mm} with a \unit[2.0]{mm} groove centered on the top side without a TiO$_2$ coating. 

The core of these extruded scintillation bars consists of a compound of Dow Styron 663W polystyrene as the base material \cite{scint_fermilab}. The attenuation length of the extruded scintillation bars is \unit[55$\,\pm\,$5]{mm} for the fast component and approximately \unit[24]{cm} for the slow component. Therefore, the light must be carried to the PMT using an optical fiber.

\subsection{Optical fibers}

To build the AMIGA MCs, the WLS optical fiber is glued in the groove with an optical cement that matches the refractive index of the fiber and the scintillator. Then, the uncoated groove, with the optical fiber inside, is covered with a reflective aluminum foil to avoid photon losses (fig. \ref{centellador}).

The fibers used in the AMIGA modules are the Saint-Gobain BCF-99-29AMC multi-clad fibers. They have the same dopant as the BCF-92 fibers (the standard catalog model by Saint-Gobain) but were ordered to have twice the concentration. The fibers are glued to the scintillators with BC-600 optical cement (clear epoxy resin) recommended by Saint-Gobain. Tests performed by the MINOS experiment \cite{minos} have shown that minimal effects from the yellowing of the scintillators and the optical cement are expected for the working conditions of the AMIGA MCs.

\section{Main sources of muon counting uncertainties}

\subsection{Muon pile-up}\label{decay_time}
   
\begin{figure}[bp]
\centering
\includegraphics[width=.47\textwidth]{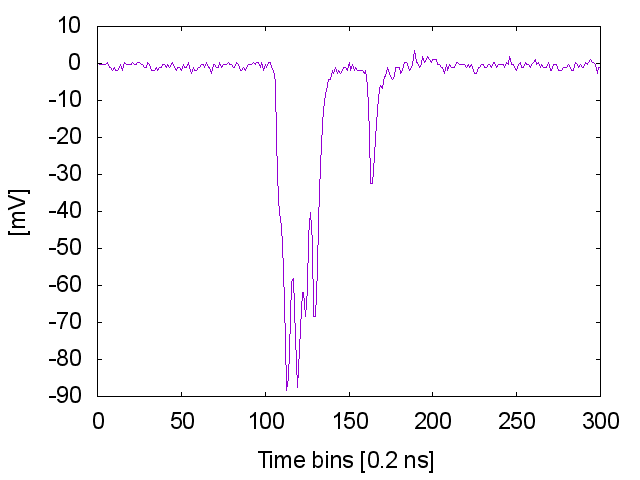}
\caption{Example of a real H8804-200MOD PMT anode output corresponding to $\sim9$ SPEs. This was generated by a muon passing through the scintillator (with an oscilloscope configured with \unit[50]{$\Omega$} input impedance, given that the mean $\text{SPE}_{peak}$ for this pixel of the PMT is \unit[$\sim26$]{mV}).}
\label{fig:resp_temporal}
\end{figure}

The convolution of fiber and scintillator decay times results in a pulse structure similar to the one shown in fig. \ref{fig:resp_temporal}, which corresponds to $\sim$9 SPEs. Greater light yields translate into higher numbers of SPEs, which in turn result in greater probabilities of wider digital muon traces (for the same discrimination threshold). As seen in the figure, the muon-signal width at 1/3 of the SPE level is $\sim$\unit[13]{ns} including the last isolated SPE. The mean number of SPEs in a module was found to be less than 25 even for situations where muons hit a point on a scintillation bar with the highest light output. 

Basically, the wider the pulse width the more over-counting for a fixed time window in which a single muon may be counted. Therefore, the mechanical design of the module considers the usage of fast optical fibers. Additionally, the fiber end opposite the PMT is painted black and clipped at 45$^{\circ}$, in order to reduce the photon reflections, which may result in delayed SPE pulses and thus larger signal widths. Painting the fiber end results in a reduction of the SPE yield of approximately one SPE (compared with no painting) at the farthest point of the scintillation bar from the PMT, and in no appreciable reduction for the closest point.

\begin{figure}[tp]
\centering
\includegraphics[width=.49\textwidth]{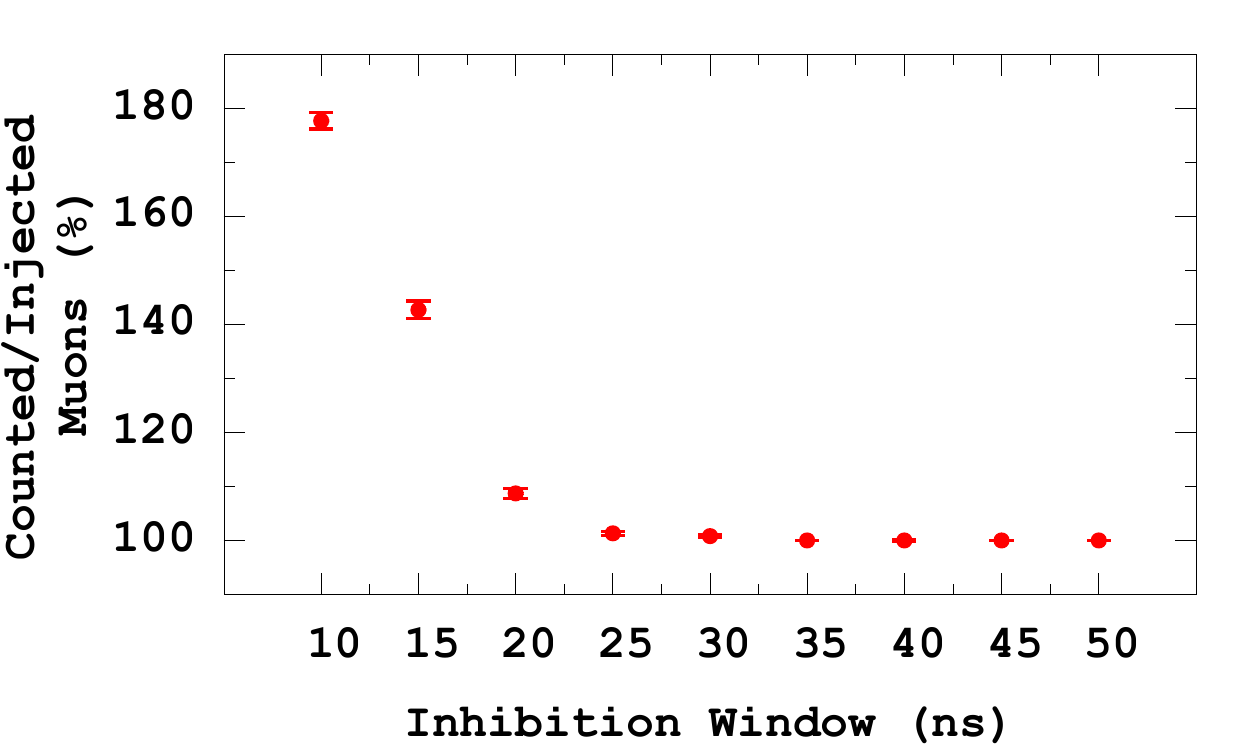}
\caption{Muon over-counting as a function of inhibition window width (simulated).}
\label{fig:inhibition_window}
\end{figure}

A simulation was performed to decide on the inhibition time window duration (fig. \ref{fig:inhibition_window}) in which a single muon may be counted. One thousand muons were injected \unit[1]{m} away from the PMT, i.e., the point of the scintillation bar closest to the PMT. A \unit[200]{ns} time span was used in the search for muon signals using a bit-pattern counting algorithm (see \cite{icrc2011:brian} for details about patterns and counting strategies in addition to their effects on counting efficiency). Starting at \unit[10]{ns}, this time span was divided in time windows increasing in \unit[5]{ns} steps. Over-counting can occur when an injected muon deposits a signal in more than one of these inhibition windows. As shown in fig. \ref{fig:inhibition_window}, 80\% over-counting would result with a \unit[10]{ns} inhibition window, 9\% for \unit[20]{ns}, and 1.3\% for \unit[25]{ns}. Virtually no over-counting results with larger inhibition windows. 

The inhibition time window is set at \unit[25]{ns} by default, but may be adjusted in \unit[3.125]{ns} steps since the reconstruction of muon numbers is performed during off-line analysis.

Muons incident on the same scintillation bar with a time separation less than the width of a time window cannot be resolved individually. Nevertheless, the number of muons in the detector can still be inferred from the number of bars with signal. Using a model that considers only the detector segmentation, it was estimated that up to approximately 700 muons could be reconstructed in a MC with a statistical uncertainty of ~10\% \cite{paper_diego}. The maximum number of muons that may be counted is limited by other factors, e.g., PMT cross-talk and inclined muons passing through more than one bar. As such, a \unit[30]{m$^2$} muon detector partitioned into 192 segments and with the inhibition time window set to \unit[25]{ns} is considered to be a good trade-off between cost and scientific output.

Even with a 10\% decrease in detection area, the uncertainties do not rise considerably, thus permitting the malfunctioning of up to 6 channels. Such malfunctions could arise from module construction, damage while transporting and/or deployment, defective PMT pixels, or defective electronics channels. 

\subsection{SPE yield and PMT cross-talk}\label{sec:light-yield}
The muon counting efficiency depends on the SPE yield of the scintillator/fiber/PMT system in addition to the counting algorithm implemented in the reconstruction. For a segmented counter there is a trade-off between muon identification and over-counting due to the PMT XT (cross-talk) between channels. The XT has been measured to be less than 1\% (charge relationship) between adjacent pixels and negligible for diagonally neighboring pixels \cite{icrc_pmts, agus_mux}. Nonetheless, the strategy adopted for muon identification is a signal where at least two non-overlapping SPE pulses are above threshold. This is because the probability of having two cross-talk SPEs (non-overlapping) in the same neighboring pixel is negligible. Note that the counting strategy can be re-adapted at the time of analysis within the software to make it more thorough and to include additional timing and adjacency conditions.

\begin{figure}[tbp]
\centering
\includegraphics[width=.85\textwidth]{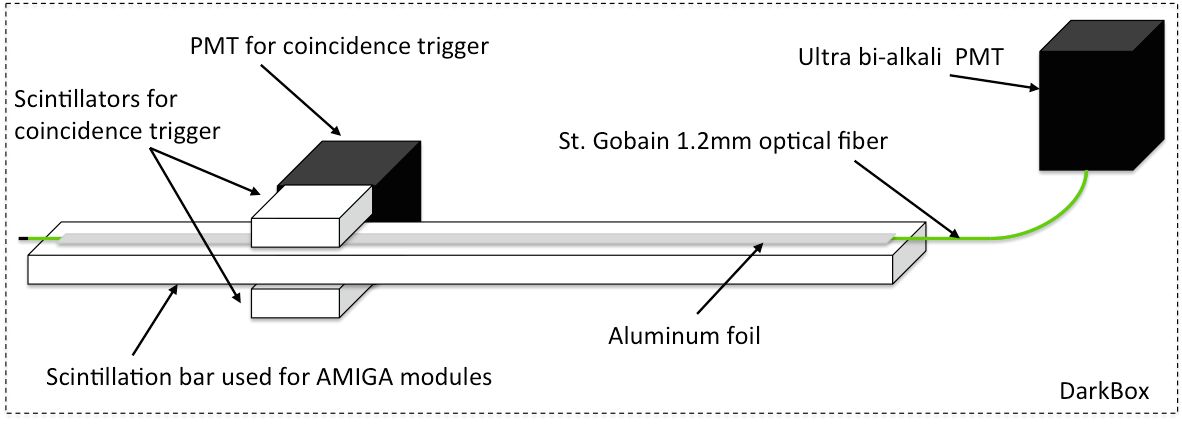}
\caption{Setup to measure the number of SPEs produced per muon.  Two \unit[4]{cm}$\,\times\,$\unit[4]{cm}$\,\times\,$\unit[1]{cm} scintillator bars and a regular multi-anode PMT were used to generate a coincidence trigger. A \unit[5]{m} Saint-Gobain fiber was glued within the groove of a \unit[4]{m} scintillation bar and curved-up at one end to optically connect it to a PMT.}
\label{fig:esquema-coincidencia}
\end{figure}

The SPE yield was also experimentally measured in a dark box with the setup shown in fig. \ref{fig:esquema-coincidencia}. Evaluation of fibers from different manufacturers (Saint-Gobain and Kuraray) and different fiber characteristics (e.g., different diameters, rounded and squared fibers) was performed to find an optical fiber with suitable mechanical characteristics that help to reduce counting uncertainties (e.g. fiber diameter, light-output). Given its narrow signal structure yet still sufficiently high light yield, the \unit[1.2]{mm} BCF-99-29AMC Saint-Gobain rounded fiber was chosen. The setup was designed to be as similar as possible to the channel of a module with the longest optical fiber, i.e., it has an extra fiber length of $\sim$\unit[1]{m}. The minimum radius of fiber curvature (produces photon losses) used in the module design to prevent damage is \unit[15]{cm}, which was also included in the experimental setup. The number of SPEs produced at the PMT photocathode by single muons was measured after calibration of the PMT. The PMT gain and its stability were monitored through the analysis of single pulses from the suspected SPE pulses. 

\begin{figure}[tp]
\centering
\includegraphics[width=.7\textwidth]{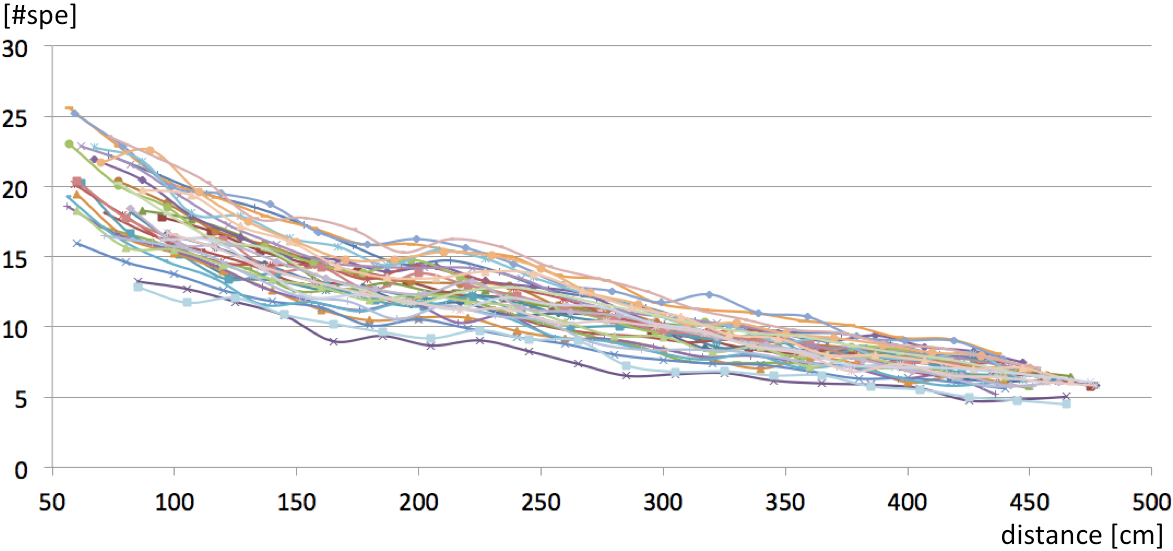}
\caption{Scanning results for PMT SPE yield vs. fiber distance for 32 channels of a module. The scanner with a radioactive source was calibrated with background muons \cite{paper_scanner}. Distance corresponds to the scanned fiber position in relation to the PMT. Minimum/maximum distances are not constant between fibers since they differ in their routing to the PMT. Regardless, the first and last points of each curve correspond to the beginning and end of the associated scintillation bar (including a safety margin at both ends).}
\label{fig:scanner}
\end{figure}

All the modules undergo quality-control testing before transportation to the Observatory. This is done by measuring the light-attenuation curves of the fibers with a \unit[5]{mCu} radioactive $^{137}$Cs source moved by an $x-y$ scanning system designed for AMIGA. The scanner allows for rapid module characterization by measuring a current which can be normalized to number of SPEs through a calibration with background muons (see fig. \ref{fig:scanner} for results). The setup for this calibration is similar to that of fig. \ref{fig:esquema-coincidencia} but with a different muon hodoscope that allows its performance on a module instead of a single scintillation bar. The scanner's operation with different PMTs to take into account variations in quantum efficiency produced similar results.

\subsection{Other counting uncertainties}
Counting uncertainties derive from multiple sources including the detector modules, the PMTs, electronics noise, counting algorithms, shower sampling resolution, shower-to-shower fluctuations, and electromagnetic punch-through. All of these sources are included in simulations of different cosmic-ray primaries. In turn, these simulations are used to calculate the so-called merit factor, which is a measure of how well it is possible to discriminate between cosmic rays of different composition. The merit factor for differentiation between proton and iron is defined as
\begin{equation}
f_{ \text{p},\text{Fe} } = \dfrac{ \left| \langle N_{\mu,\text{Fe}}^{opt} \rangle - \langle N_{\mu,\text{p}}^{opt} \rangle \right| }{ \sqrt{ \sigma^2_{\mu,\text{Fe}} + \sigma^2_{\mu,\text{p}} } }\text{ },
\end{equation}
where $N^{opt}$ is the number of muons at a given optimum reference distance and $\sigma$ is the standard deviation therein. The merit factor is estimated to be 2.1 and 1.7 for a primary with an arrival angle of 0$^\circ$ and 38$^\circ$, respectively (at an energy of \unit[10$^{18}$]{eV} and considering \unit[30]{m$^2$} MCs separated by \unit[750]{m} and buried \unit[1.3]{m} underground). Further discussion of the merit factor is not within the scope of this paper and will be reserved for elsewhere. 

%
%
%

\section{Mechanical design}
\label{sec:mechanical-design}

\subsection{Module casing}

\begin{figure}[tbp]
\centering
\includegraphics[width=.46\textwidth]{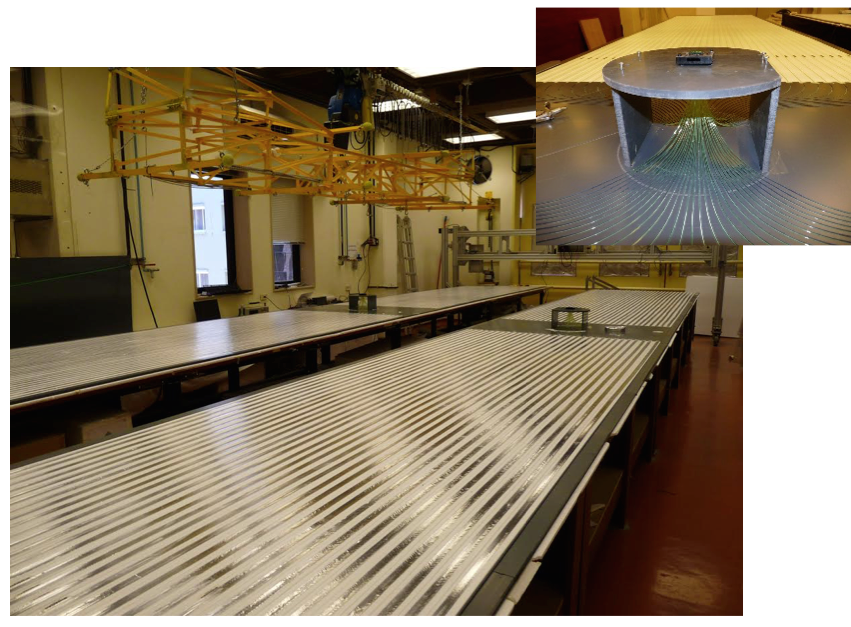}
\includegraphics[width=.53\textwidth]{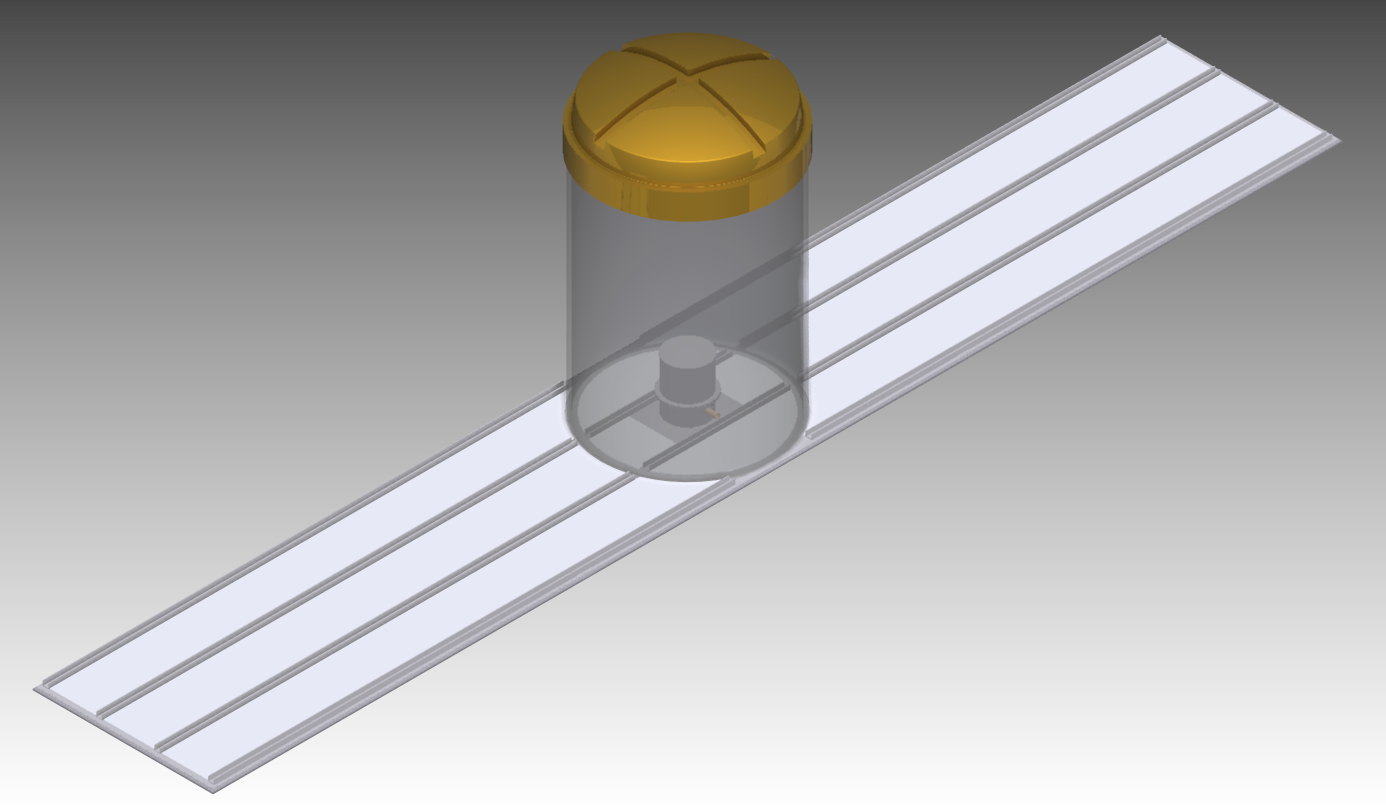}
\caption{Left: picture of two \unit[10]{m$^2$} modules being manufactured at the same time in the ITeDA (Instituto de Tecnolog\'ia en Detecci\'on y Astropart\'iculas) facilities. The 64 scintillation bars are divided into two groups of 32 bars each, with all optical fibers curving up into the centrally located optical coupling device for the PMT. The scanner using a radioactive source can be seen at the back of the facility. Right: diagram of the PVC casing of the \unit[10]{m$^2$} scintillator module and centrally located electronics dome.}
\label{fig:foto_armando_mc}
\end{figure}

To follow the requirements mentioned in the previous sections, a decision was made to split the module in two halves with 32 bars on each side of a centrally located PMT (fig. \ref{fig:foto_armando_mc}). This design has the following advantages:
\begin{itemize}
\item It allows for the use of standard trucks for transportation thereby reducing the project costs.
\item It produces less optical fiber wastage due to the reduced distance between the farthest scintillation bar and the PMT.
\item It facilitates easy access to the modules for maintenance after deployment. The module itself can be used as an access platform for the technicians. This is possible because the PMT sits in the middle of the module and not beside it.
\item It is more robust and permits a simpler support scheme for the electronics.
\item It allows for use of commercial \unit[2]{mm} PVC plates for the casing with a standard size of \unit[1.5]{m}$\,\times\,$\unit[3]{m}, thereby considerably reducing the costs and the number of plates to be sealed to fabricate the casing.
\item It facilitates deployment of the fibers in the scintillator grooves in two groups of 32 during assembly. This is more convenient for the technicians to work comfortably on every part of the module. 
\end{itemize} 

\subsection{Optical coupling and alignment system}
\label{sec:optical_connector}

\begin{figure}[bp]
\centering
\includegraphics[width=.77\textwidth]{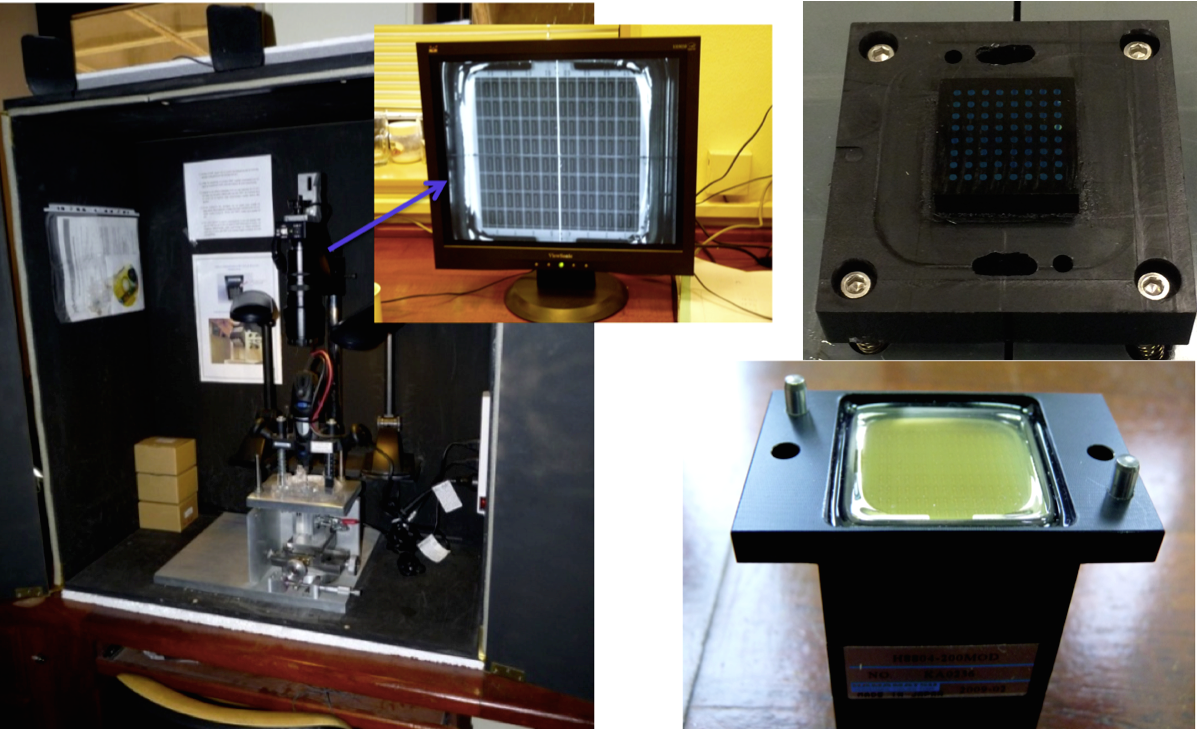}
\caption{Alignment system for coupling of PMT and optical fiber. Left: the X-Y-W (W for rotation) alignment bench and the high-resolution camera. Right top: a module's optical connector with the 64 fibers glued. Right bottom: the PMT with metal pins inserted into the holes drilled in the casing.}
\label{fig:alineacion}
\end{figure}

The optical fibers and the PMT are joined by an optical connector (fig. \ref{fig:alineacion}, top right) made out of black POM (black Polyoxymethylene). POM was chosen for its good machining properties, which allow fast and precise milling without deformation or material burning. Black color was chosen to help to reduce optical XT among the PMT pixels due to reflections in the PMT glass-photocathode boundary. Note that since the connector is the same for all modules, there is no need to replace it if a PMT requires swapping.

During the assembly process, the 64 optical fibers are threaded and glued inside the drilled holes of the optical connector. Then a fly-cutter milling machine is used to simultaneously cut and polish all the fiber ends and the connector's front face. As a result, the fibers are left flush with the optical connector, which lies flat against  the PMT glass reducing the optical XT between neighbouring pixels.  

\begin{figure}[bp]
\centering
\includegraphics[width=.7\textwidth]{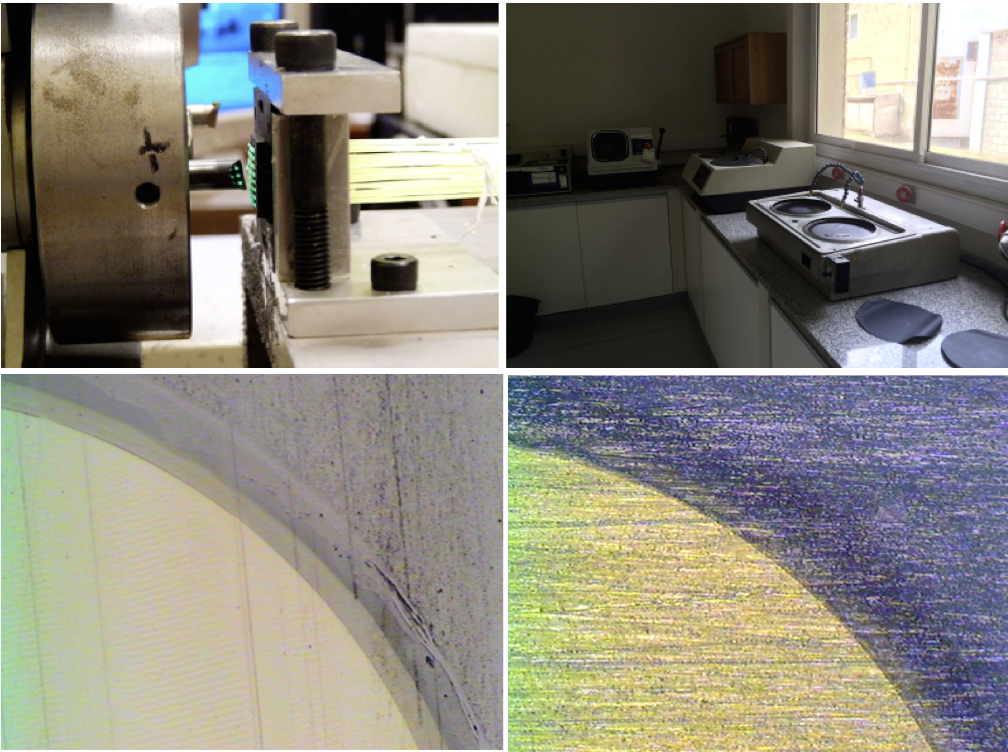}
\caption{Results of using a fly-cutter with diamond tools (left) versus a fast fiber polishing with sand paper (right) up to \#1500 (\unit[1500]{grits cm$^{-2}$}). The pictures at the bottom correspond to a quarter of the optical fiber taken with an Olympus bx60m microscope (10x brightfield). The stripes left by the fly-cutter are the very subtle nearly vertical lines (bottom left picture).}
\label{fig:flycutter}
\end{figure}

Fly-cutting was chosen over polishing with regular sandpaper because the former is faster and reduces the risk of leaving residuals inlaid in the polished face of the soft plastic fibers (fig. \ref{fig:flycutter}), that may increase the optical XT and/or reduce the SPE yield. Furthermore, sandpaper polishing is typically performed as a wet process, which is not easily applicable to the mechanical design of the modules and their assembly procedure. A comparison with a rapid polishing is shown in the figure. Detailed polishing is precluded by the mass-production timeline. Finally, the optical coupling is then enhanced with Saint-Gobain BC-630 optical grease to better match the refractive indexes.

The PMTs (which were also used by the Opera experiment \cite{Opera}) are delivered by Hamamatsu in their POM casing ready for the alignment procedure (fig. \ref{fig:alineacion}) consisting of the insertion of two metal pins aligned with the first dynode of the pixels using an X-Y-W positioning system. These metal pins then match with the alignment holes in the optical coupler.

\subsection{Module handling and transportation}

\begin{figure}[tbp]
\centering
\includegraphics[width=.49\textwidth]{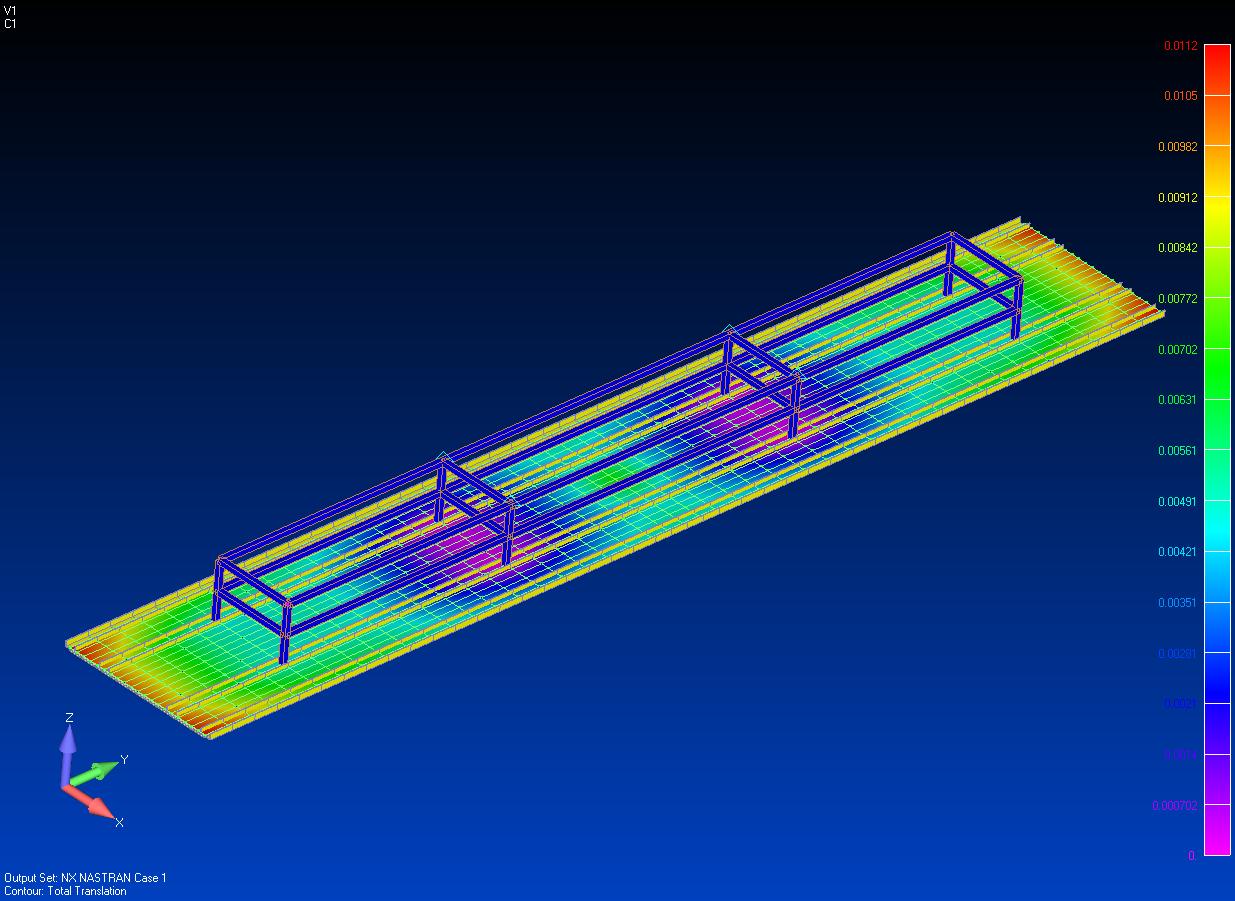}
\includegraphics[width=.48\textwidth]{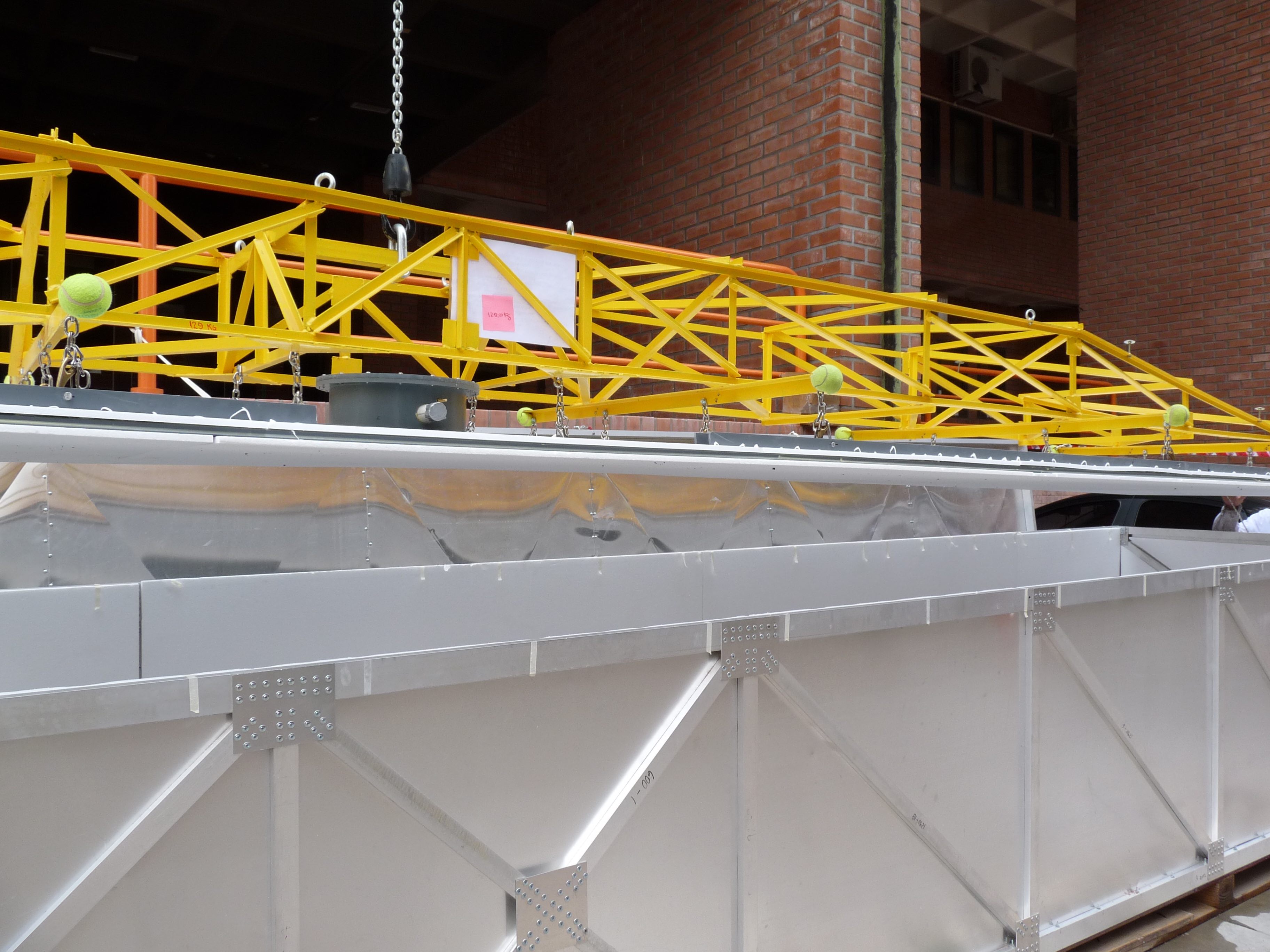}
\caption{Left: finite-element simulation of module handling with a simplified hanger structure \cite{Terlinsky:08} showing the suggested attachment points. Right: a \unit[10]{m$^2$} module being lifted. The detector is being extracted from the laboratory and placed inside the transportation container.}
\label{fig:elementosfinitos}
\end{figure}

After performing the corresponding finite-element simulations (fig. \ref{fig:elementosfinitos}, left), a decision was made to glue four ``U" PVC profiles along the length of the module to enhance its structural integrity and provide grabbing points for handling (only for the AMIGA UC module). Modules are lifted and handled with a steel-hanger structure that attaches to the modules along their top surfaces (fig. \ref{fig:elementosfinitos}, right).

A container was built to allow for the simultaneous transport of up to four modules, which permits it to be lifted by standard fork-lifts (each module weighs approximately \unit[300]{kg}) to reduce logistical costs. 

\section{Construction and testing}

\subsection{Manufacturing}
The gluing of the fibers on the scintillators is labor and skill intensive. The process must be administered as uniformly as possible, and careful inspection is necessary to ensure good optical coupling along the fiber length. Experience has shown that this process is difficult to automate. 

The rest of the assembly consists of gluing the rest of the PVC parts (e.g., manifold plate, optical coupler and its holder) and the scintillation bars in the PVC casing with Teroson MS 939 adhesive (fig. \ref{fig:pegamento}, left). The optical connector is then attached to the central holder and the fibers are glued into the bars' grooves as previously explained. The top PVC plates are also glued to the scintillation bars with Teroson MS 939 adhesive to close the module casing. The borders of the modules are sealed with Plexus MA310, which is a solid 2-component glue that prevents ground water from entering.

\subsection{Mechanical and quality assurance tests}
\begin{figure}[tbp]
\centering
\includegraphics[width=.49\textwidth]{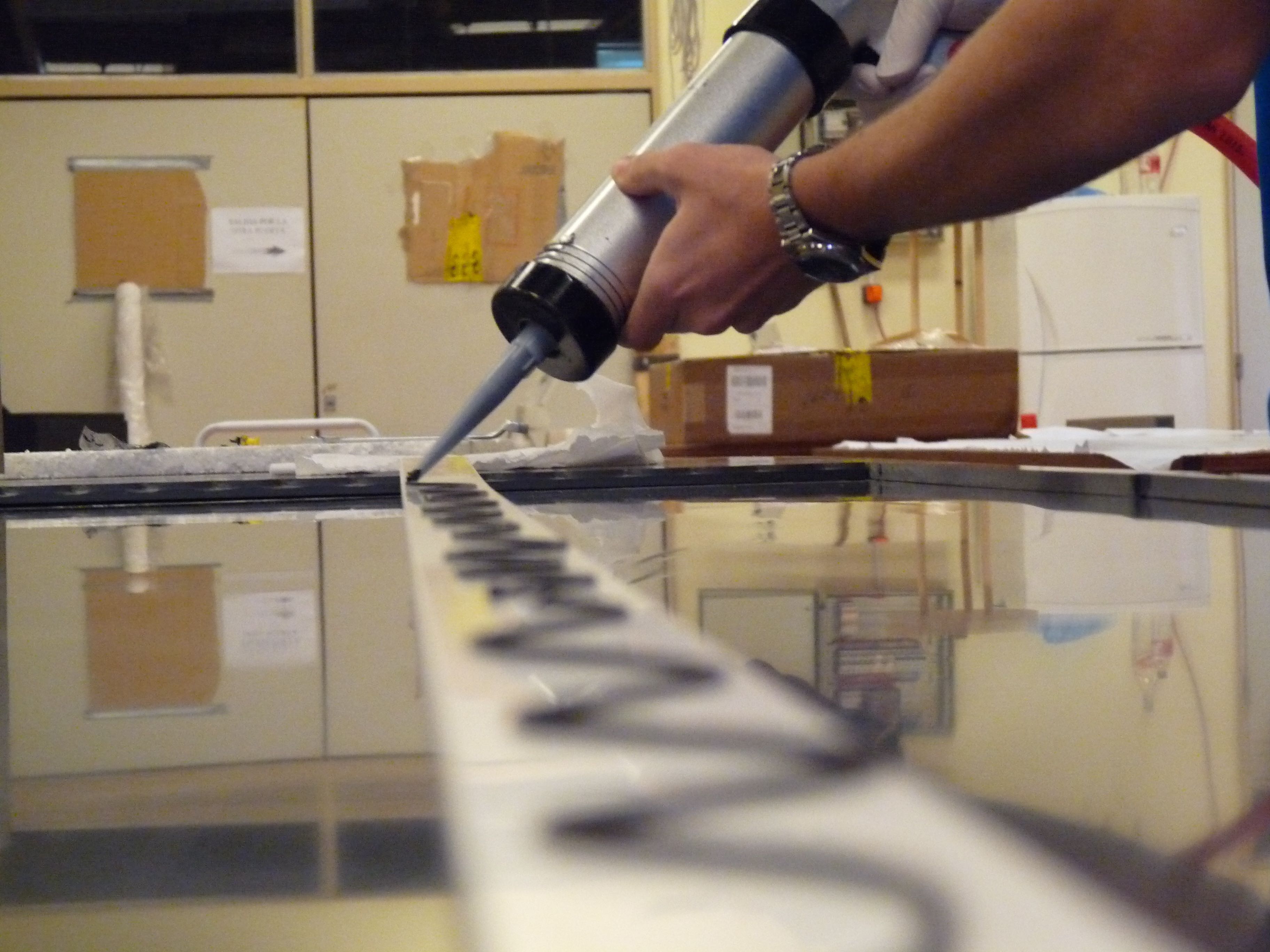}
\includegraphics[width=.49\textwidth]{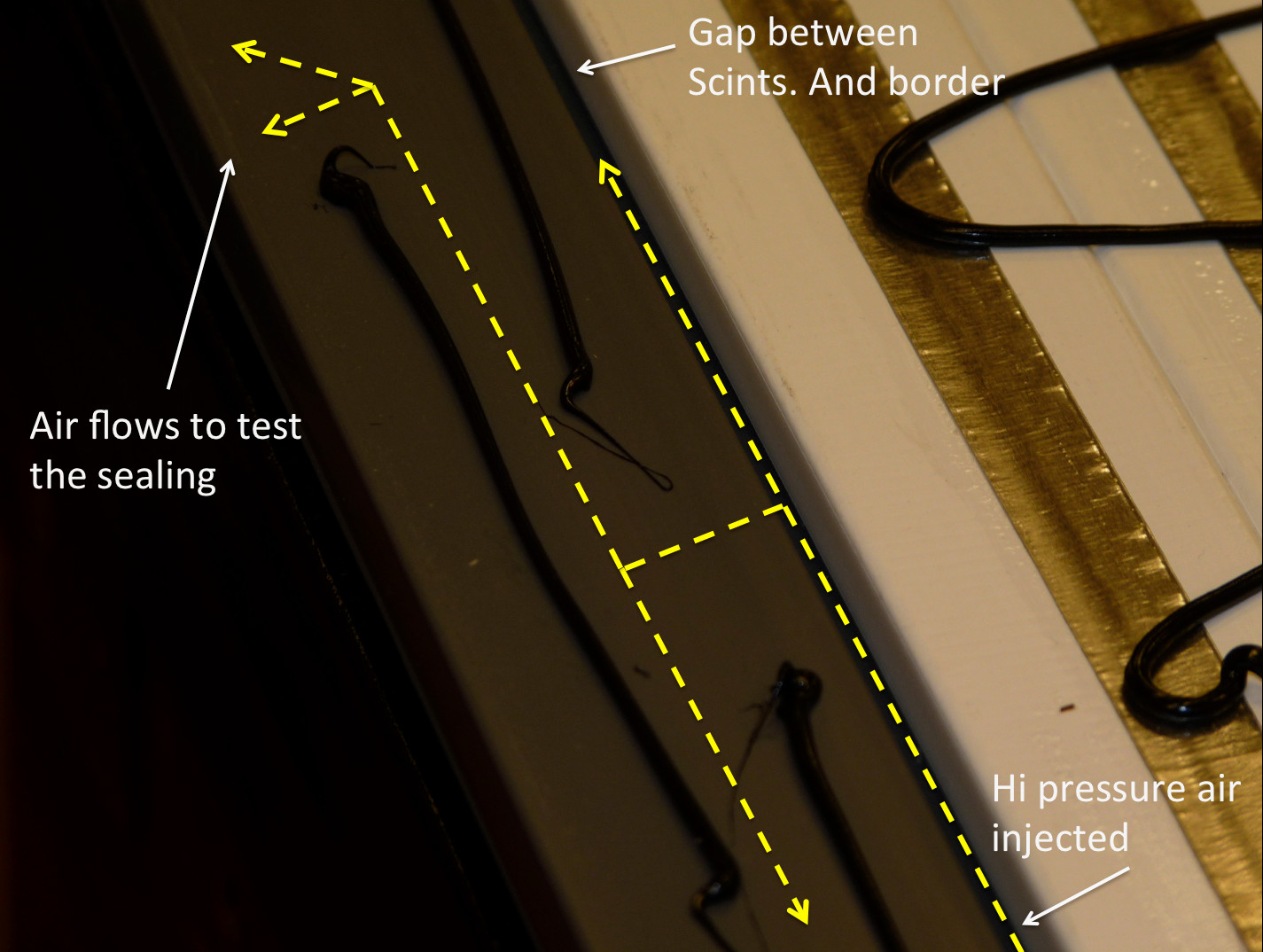}
\caption{Left: process of gluing the scintillation bars into the PVC casing. Right: process of gluing the frame PVC bars to the PVC sheets of the casing, which allows for hermeticity testing.}
\label{fig:pegamento}
\end{figure}

Pneumatic leak tests are administered after sealing the casing (fig. \ref{fig:pegamento}, left). The gluing process for the Teroson MS 939 adhesive is performed in a way that allows for the passage of air (fig. \ref{fig:pegamento}, right) throughout a module to test its hermeticity. After performing this test, the port used to pressurize the module is left unsealed in order to compensate for changes in atmospheric pressure during travel. This venting hole is filled with Plexus glue once the module arrives at its destination.

One final test is performed to identify defective channels and to inspect their light output. This is done using the scanning system discussed in section \ref{sec:light-yield}. All of the modules in the UC including the twin detectors were tested. The rate of defective channels was found to be approximately a third of a percent. Failures were due to either fiber manufacturing defects or incorrect fiber/scintillator gluing.

\section{Deployment}

Prior to deployment, a back-hoe digger excavates the pits in which modules are to be placed. A levelled fine sand bed is prepared at the bottom of the pit to provide a flat platform for the modules. The modules are then inserted into the pits with the same back hoe and covered with approximately \unit[20]{cm} of local sand to provide additional protection during the refilling of the pits 

\subsection{Ground shielding uniformity}
The AMIGA UC modules are buried approximately \unit[2.25]{m} underground, which amounts to approximately \unit[540]{g cm$^{-2}$} of vertical mass (a total of about 60\% more than the atmosphere at the level of the Auger Observatory, namely, 870 g cm$^{-2}$), to provide shielding from electromagnetic particles. Two modules have also been deployed to examine whether rates of electron punch through are tolerable at a depth of \unit[1.3]{m} (fig. \ref{mapaAugerAMIGA}, right).

\begin{figure}[tbp]
\centering
\includegraphics[width=.59\textwidth]{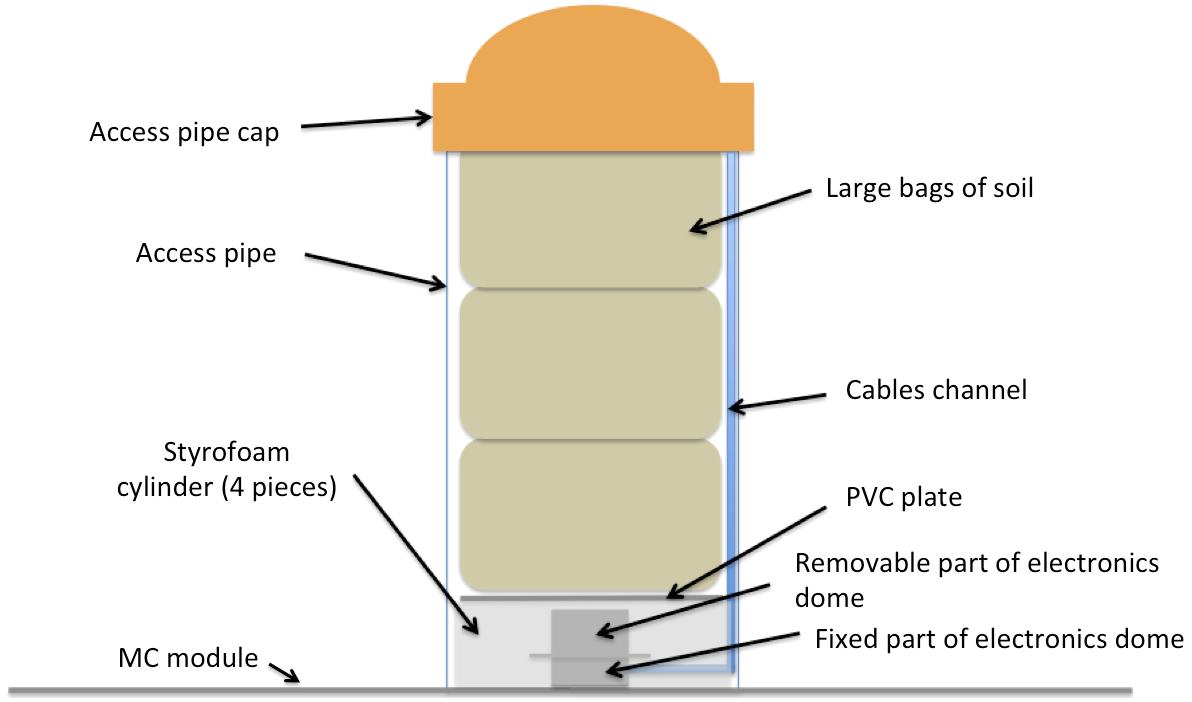}
\includegraphics[width=.39\textwidth]{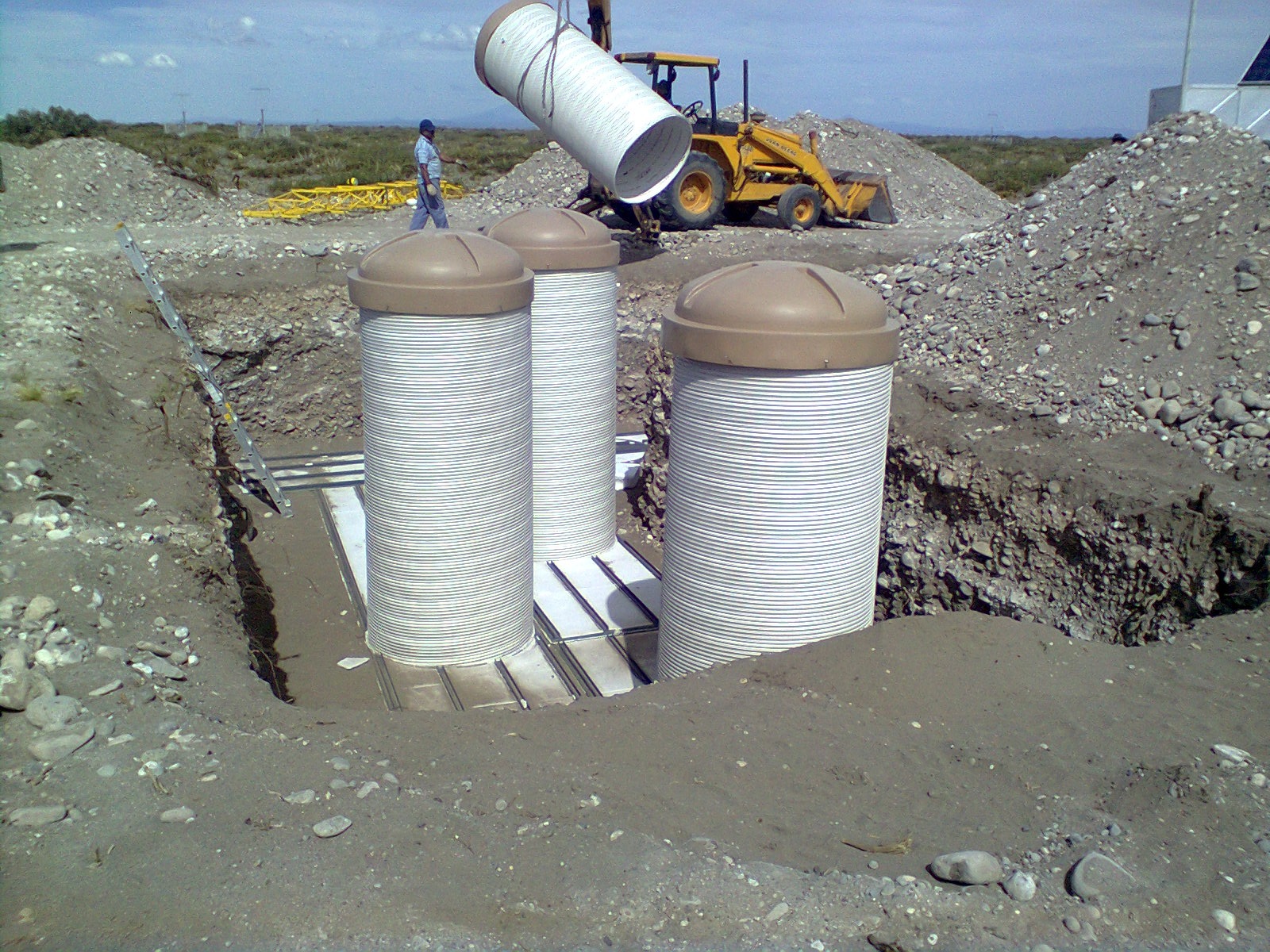}
\caption{Left: Schematic of the module-access tube used for electronics maintenance. Right: installation of an access tube at deployment time. The access tube is sealed to the module for water tightness.}
\label{fig:accesstube}
\end{figure}

Since the module is buried, its mechanical design provides a way to access both the electronics and the PMT for maintenance purposes. To do this, an access tube that is filled with large commercial grade bags of local soil after maintenance to ensure uniform shielding (fig. \ref{fig:accesstube}, left).  Once filled, the access tube is covered with a lid designed to resist sun exposure, large animals, and high-speed winds.

\subsection{Layout}
\begin{figure}[tbp]
\centering
\includegraphics[width=.77\textwidth]{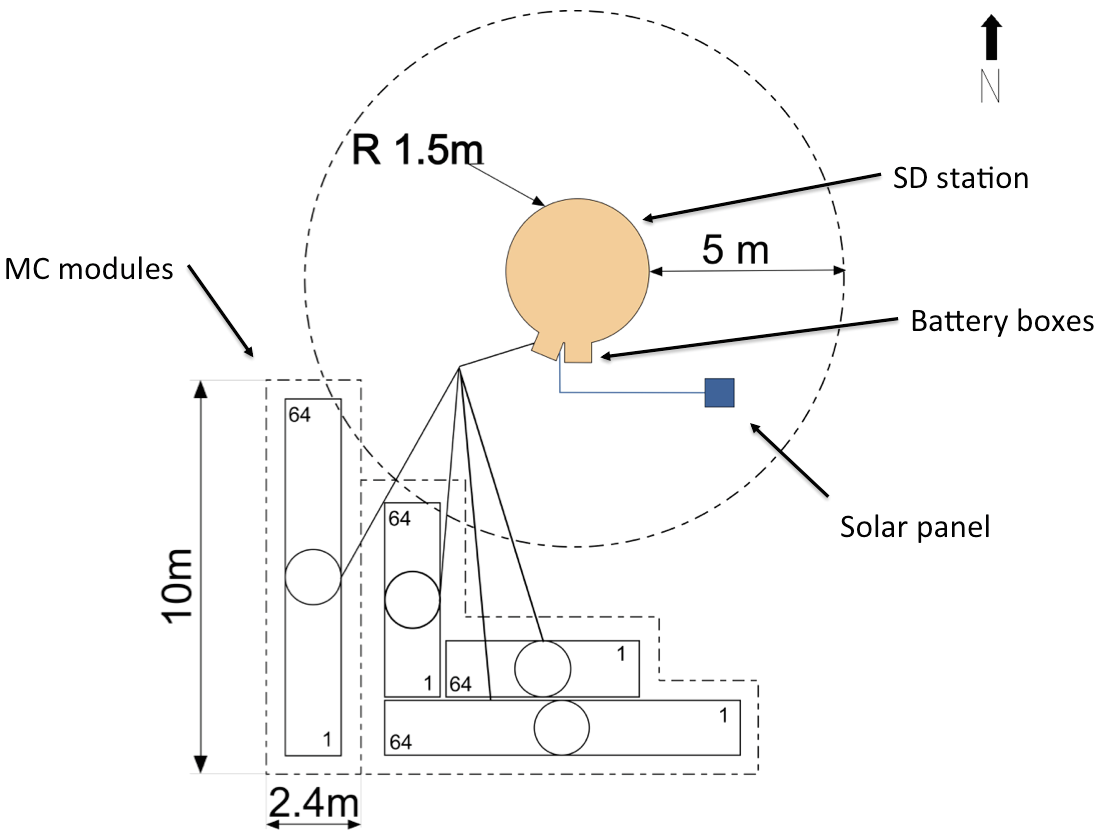}
\caption{Module layout of an MC in the AMIGA UC. The particular \unit[10]{m$^2$} module at the left here is separated from the rest because it was deployed at an earlier point in time with respect to the other three. There is a horizontal separation of approximately \unit[5]{m} between the SD station and the modules to improve the shielding uniformity. The excavation area is marked with a dashed line.}
\label{fig:layout}
\end{figure}

The four modules of each UC MC are arranged in an ``L" shape (fig. \ref{fig:layout}) in an effort to minimize possible systematics arising from highly-inclined muons.

The modules are deployed with a horizontal separation of approximately \unit[5]{m} from the SD station in an effort to guarantee shielding uniformity independently of the shower arrival direction. On the other hand, larger separations are avoided in order to prevent excessive cabling for power and data.

\section{Module data acquisition}\label{sec:data_acq}

The electronics of the MCs are split in two: the "underground electronics" installed in each buried module and additional "surface electronics" at the SD station common to all of the modules of the MC. Both are powered by solar panels (fig. \ref{fig:estacion}). 

The surface electronics are comprised of the interface with the SD electronics (trigger and muon data transfer), the wireless communication to the Observatory CDAS (Central Data Acquisition System), the network switch to communicate with the modules of the MC, and the power regulator of the photovoltaic system. The underground electronics include the PMT, the analog front-end, a digital board with an FPGA (Field Programmable Gate Array) and memories, and a micro-controller board for interface, data transmission and slow control. 

The PMT is a 64-pixel Hamamatsu H8804-200MOD, which is a H7546 device with a different casing and a high quantum efficiency ultra bi-alkaline photocathode, i.e., the quantum efficiency is still approximately 21\% at the optical fiber emission peak. The last dynode output is also available and is common to all channels in these PMTs, which facilitates signal-charge integration of the whole module.

The analog front-end holds 64 pre-amplifiers and discriminators which are remotely set to an adjustable fraction of the average SPE amplitude of each PMT pixel. In this way, the PMT pulses are converted into a train of 0s and 1s, which correspond to the respective absence or presence of a signal above the mentioned threshold.  One bit per channel is saved in the front-end memory forming a 64-bit character word per time bin. This conversion is performed in \unit[3.125]{ns} time bins in the FPGA (see \cite{jinst:oscar} for details about the electronics kit). Basically, the memory consists of two circular buffers that store 2048 bins of 64 bits. Following an SD trigger these bit trains are stored. They are then recovered and transmitted upon a request from the CDAS.

\section{AMIGA sample event}
\begin{figure}[tbp]
\centering
  \includegraphics[width=.8\textwidth]{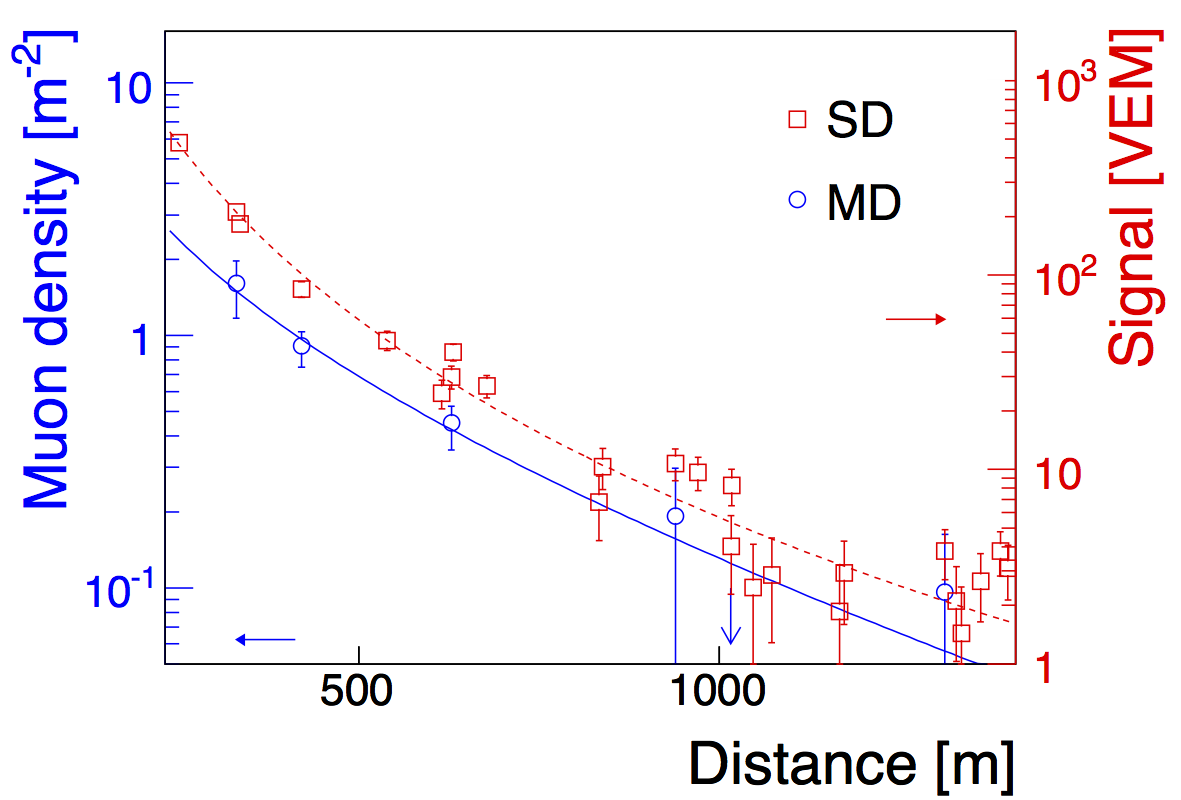}
  \includegraphics[width=.99\textwidth]{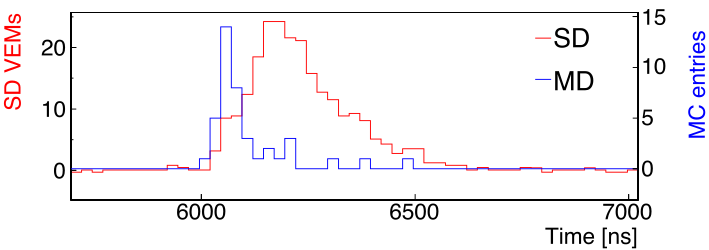}
  \caption{First analysis of a sample event detected in coincidence by the SD and AMIGA. Top: Lateral distribution functions fitted from data recorded by each detector. Bottom: muon arrival times for the MC at \unit[420]{m} from the shower core.}
  \label{evento}
 \end{figure}

The MC modules are currently acquiring and transmitting data to the CDAS. A first analysis of an event demonstrating the capability of the MCs to count muons and measure their arrival times is shown in fig. \ref{evento}. This event corresponds to a $8\times10^{17}$ eV shower arriving with a zenith angle of $19^\circ$ where the shower core landed within the UC. The event occurred while the UC was under construction and six MCs participated in the event. At the time four MCs had a detection area of \unit[10]{m$^2$}, one \unit[45]{m$^{2}$}, and one \unit[60]{m$^2$}. The number of reconstructed muons per MC was 16, 2, 1, 0, 43 and 28 respectively. The MC with zero muons was also considered in the fit to the lateral distribution function of muons (blue downwards arrow). This event shows the quality performance of the module design and the data reconstruction used for muon counting.

\section{Production mechanical design}

Expertise gained from the UC informed the development of some design modifications for the AMIGA production phase. These modifications targeted cost reduction, faster mass production, and streamlined deployment. The main upgrades already implemented in the mechanical design are: PVC folding to produce the module casing, modules lifting with a hanger with vacuum suction cups, \unit[30]{cm} diameter access tube for maintenance, and better deployment procedure to avoid usage of styrofoam. See appendix \ref{sec:details} for additional technical information if desired.

\section{Conclusions}

The design of the UC muon counters has been shown to function successfully within expectations. The mechanical design of the modules has shown quality results regarding mechanical and functional stability, performance of the couplings with PMTs, light and water tightness, and temperature stability.

No major mechanical damage or module losses were suffered during the construction and deployment of the 38 modules of the UC, the twin detectors, and other prototypes.

First longitudinal and lateral muon profiles have been reconstructed and are currently under further analysis. The stable and quality performance of the AMIGA muon counters is reflected in the analysis of events recorded thus far.
Important advances in the engineering of the muon counters' logistics (i.e., fabrication procedures, transportation, and deployment techniques) have been achieved, which have allowed for much progress during the re-design phase for module mass-production. Some upgrades to the mechanical design for production were described here. The goals of these upgrades are primarily to increase production rate and to reduce the cost per detector module for fabrication/production and post-deployment maintenance.

\appendix
\section{Additional technical information}\label{sec:details}

\subsection{Module dimensions}
The UC includes modules with different detection areas, \unit[5]{m$^2$} and \unit[10]{m$^2$}, both of which have 64 channels, which results in a factor of 2 difference in segmentation area. The \unit[5]{m$^2$} and the \unit[10]{m$^2$} modules have external dimensions of approximately \unit[5]{m}$\,\times\,$\unit[1.4]{m} and \unit[9]{m}$\,\times\,$\unit[1.4]{m}, respectively. Both are built with the same technique and specifications except for minor mechanical details.

\subsection{Scintillators}
The dopants follow the proportions: 1\% PPO [2,5-diphenyloxazole] and 0.03\% POPOP [1,4-bis(5-phenyloxazole-2-yl)benzene], by weight of base material and with concentration variations of no more than $\pm10\%$. These dopant components result in a blue-emitting scintillator with an emission maximum of approximately \unit[420]{nm}. The outer layer is extruded in a second extrusion machine and consists of a compound of clear polystyrene with TiO$_2$ in a concentration of $15\%$ by weight of the base material (with variations of no more than $\pm10\%$). The scintillation light yield uniformity is quality-controlled when the bars are manufactured to be constant to $\pm5\%$.

\subsection{Optical fibers}
The dependence of the muon arrival-time distribution on the distance of a MC from the shower core allows for reconstruction of the muon production depth. For this reason and the problem of pile-up, the temporal characteristics of the detector modules are of great importance. As mentioned before, a fast Saint-Gobain fiber is used, which complements the scintillator decay time of \unit[3.7$\,\pm\,$0.5]{ns} (measurement performed by the manufacturer).

The BCF-99-29AMC multi-clad fibers (same as the fast BCF-92 but twice the dopant concentration) have a second layer of cladding that has an even lower refractive index and thus permits total internal reflection at this second boundary. The additional photons guided by multi-clad fibers increase the output signal by up to 60\% as compared to conventional single-clad fibers. The BCF-92 fibers have maximum light absorption at \unit[410]{nm} and maximum emission at \unit[485]{nm} for the absorption and emission spectra. Also, they have a decay time of \unit[2.7]{ns}, which makes them ideal for use with the scintillation bars chosen for AMIGA. The standard cladding material for Saint-Gobain fibers is PMMA (polymethylmethacrylate, C$_5$H$_8$O$_2$). It has a density of \unit[1.2]{g cm$^{-3}$} and a refractive index of 1.49. The trapping efficiency of these round fibers ranges from 3.4\% to approximately 7\% for events occurring at the fiber axis and near the core-cladding interface, respectively. 

The fiber diameter has a direct impact on the light output of the scintillator channels. Measurements taken with prototype detectors showed that both \unit[0.8]{mm} and \unit[1.2]{mm} fibers generate enough light output for the \unit[5]{m$^2$} and \unit[10]{m$^2$} modules, respectively. However, \unit[0.8]{mm} fibers are not suitable for the \unit[10]{m$^2$} modules. As such, \unit[1.2]{mm} was adopted for both module sizes for manufacturing convenience.

The scintillator-fiber transmission coefficient (BC-600 optical cement) is expected to remain constant due to the position of the fiber (glued at the very bottom of the groove) and at a working temperature measured to be lower than 20 degrees Celsius with only seasonal variations.
 
\subsection{PVC casing}

The module casing is comprised of PVC including its structural profiles and the electronics dome. It is formed by a frame (PVC bars with a section of \unit[1]{cm}$\,\times\,$\unit[4]{cm}) all along the perimeter and is closed with \unit[2]{mm} extruded PVC plates on the top and bottom. The 64 scintillation bars fit inside this casing. All the PVC pieces are custom-made (cut and/or glued) in the project facilities from industrial extruded PVC bars, sheets, and tubes (for the electronics dome). PVC was chosen because it is inexpensive, light, resistant, easy to machine, and easy to seal with thermal solder or glue. 

\begin{figure}[bp]
\centering
\includegraphics[width=.7\textwidth]{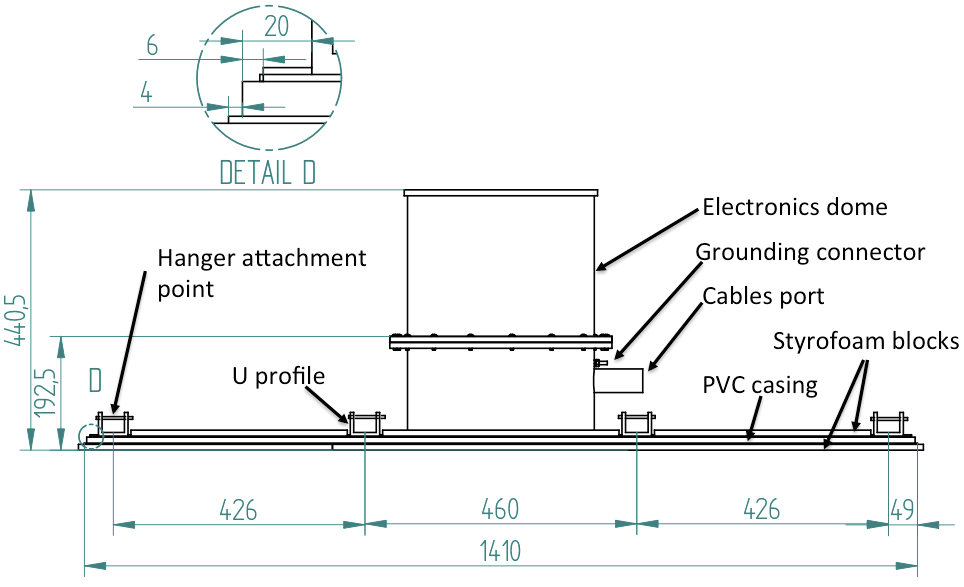}
\includegraphics[width=.29\textwidth]{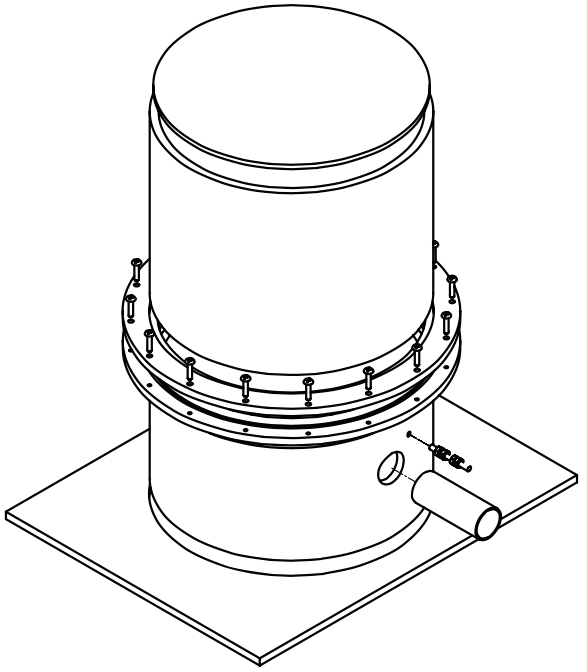}
\caption{PVC casing of the scintillator modules (measures in mm). Left: longitudinal view of the module. Styrofoam blocks glued on top and underneath the module provide mechanical protection and prevent sun exposure. Right: electronics-dome parts.}
\label{fig:plano_modulo_cortelong_domo}
\end{figure}

The manufacturing of most of the module's components (figs. \ref{fig:foto_armando_mc} and \ref{fig:plano_modulo_cortelong_domo}) is automated by the use of a computerized numerical control router milling machine. The PVC electronics dome (fig. \ref{fig:plano_modulo_cortelong_domo}) is split into two main pieces. The lower one is fixed to the PVC casing and provides an outlet for the cabling and protection for the enclosed curved fibers. The upper part is removable and allows access to the electronics in the event that replacement or maintenance is required. The pieces are sealed together with a rubber O-ring which provides water- and light-tightness.

\subsection{Optical coupling}
The fly-cutter is equipped with two tool bits with diamond inserts. In addition, this machine is supported by a portable anti-vibrating structure to reduce creation of stripes and to allow for operation close to the modules during assembly.

The optical connector is supported within the fixed part of the module's dome with $\sim$\unit[1]{cm} of floating freedom to allow for compensation of any difference with the electronics holding support and the PMT dimensions. The optical connector attaches directly to the PMT (which is fixed to the electronics) and is held in place by a couple of springs providing enough pressure to ensure good optical coupling.

The positions of the pixels (focalization electrodes and first dynodes for each of the 64 channels) of different PMTs vary by some tenth of a mm with respect to the casing. The final alignment is performed using 4 distinct alignment marks (dots) provided by Hamamatsu on each side of the PMTs. These indicators define two orthogonal lines which intersect at the PMT center. As such, a metal cross and a high-resolution camera are used to align the PMT alignment dots and the optical connector with an X-Y-W (W for rotation) system (fig. \ref{fig:alineacion}). Once alignment is complete, two metal pins are inserted into holes drilled in the PMT casing, which precisely align with the optical connector holes (previously drilled while machined).

\subsection{Transportation container}
A container was built to allow for the simultaneous transport of up to four modules, made of aluminum for its lightness, which permits it to be lifted by standard fork-lifts (each module weighs approximately \unit[300]{kg}) to reduce logistical costs. It was designed to resist deformation during elevation and placement onto a truck trailer. Moreover, this container is also used to transport modules up to the deployment site without needing to make any transfer at the Observatory facilities. A \unit[5]{cm} thick foam layer was installed at the bottom of the container in order to protect the modules against any vibrations of the truck. Low density styrofoam was inserted in between the modules to reduce vibrational effects.

\subsection{Construction, testing and deployment}
Threading the 64 fibers into the optical connector is skill intensive. For ease of process and minimization of damage risk, the optical fibers are cut at $\sim$45$^{\circ}$ to make a sharp ending for threading through the connector holes. Connector holes are drilled slightly larger than the fiber diameter to reduce instances of fibers cracking during threading but not large enough such that glue leakage occurs.

The ``bubble test" is performed by pressurization of the module with compressed air (1.02 atmospheric pressure) and applying soapy water to the sealed borders to search for bubbles (see \cite{bubble_test} for details of the standard testing method for leaks). 

During deployment, the walls of the pits are excavated approximately \unit[50]{cm} wider than the size of the modules and with a step shape in the top half meter in order to reduce risks associated with falling rocks and walls collapsing. The soil structure was, however, very firm and compact, and no collapses have been observed.

The service access for the electronics was designed from a PVC tube with a diameter of \unit[1.3]{m} (built out of a PVC strip, not extruded). The access tube is filled with large commercial grade bags of local soil after maintenance to ensure uniform shielding. A high density styrofoam structure sits between the electronics dome and these bags, in order to protect the former.

The access tubes are transported independently of the detector modules, and installed at deployment time (fig. \ref{fig:accesstube}, right). This involves sealing them to the module's upper surface with Plexus MA310 adhesive to ensure that their position remains static and that they are water tight. 

\subsection{Proposals for the production mechanical design}

The main upgrades already implemented in the mechanical design are:

\begin{itemize}
\item \textit{PVC folding:} The UC modules have PVC bars at the edge of the casing to which the top and bottom PVC sheets are glued (fig. \ref{fig:pegamento}, right). A prototype module has already been manufactured for which these bars were replaced by folding of the edges of the bottom PVC sheet to glue directly to the scintillation bars and the top PVC plate. This procedure reduces the number of parts in a module and requires approximately half of the MA310 glue sealing. Additionally, it reduces the need for leak testing, and the period during assembly for which the use of a breathing mask is necessary.
\item \textit{No U-profiles:} The production prototype module does not have PVC U-profiles. It is lifted with rubber vacuum suction cups as is done in the glass industry.
\item \textit{Small access tube:} the access tube is now constructed with a \unit[30]{cm} diameter PVC tube that permits the removal of the electronics and the PMT with a probe. This reduces the cost since the small diameter tubes are much cheaper than those of \unit[1.3]{m} diameter, and do not require the use of cranes to remove the large bags of soil. Additionally, these tubes allow for faster, easier, and safer deployment.
\item \textit{No styrofoam:} the deployment procedure will be improved to ensure no rocks are present in the sand protection layers. Additionally, the elevation hanger will provide protection from sun exposure during deployment.

\end{itemize}

\acknowledgments

The successful installation, commissioning, and operation of the Pierre Auger Observatory would not have been possible without the strong commitment and effort from the technical and administrative staff in Malarg\"{u}e. 
We are very grateful to the following agencies and organizations for financial support: 

 Comisi\'{o}n Nacional de Energ\'ia At\'{o}mica, 
 Agencia Nacional de Promoci\'{o}n Cient\'ifica y Tecnol\'{o}gica (ANPCyT), 
 Consejo Nacional de Investigaciones Cient\'ificas y T\'ecnicas (CONICET),
 Gobierno de la Provincia 
 de Mendoza, Municipalidad de Malarg\"{u}e, 
 NDM Holdings and Valle Las Le\~{n}as, in gratitude 
 for their continuing cooperation over land access, 
 Argentina; the Australian Research Council; Conselho 
 Nacional de Desenvolvimento Cient\'{\i}fico e 
 Tecnol\'{o}gico (CNPq), Financiadora de Estudos e 
 Projetos (FINEP), Funda\c{c}\~{a}o de Amparo \`{a} 
 Pesquisa do Estado de Rio de Janeiro (FAPERJ), 
 S\~{a}o Paulo Research Foundation (FAPESP) 
 Grants No. 2010/07359-6 and No. 1999/05404-3, 
 Minist\'{e}rio de Ci\^{e}ncia e Tecnologia (MCT), 
 Brazil; Grant No. MSMT-CR LG13007, No. 7AMB14AR005, 
 and the Czech Science Foundation Grant No. 14-17501S, 
 Czech Republic;  
 Centre de Calcul IN2P3/CNRS, Centre National de la 
 Recherche Scientifique (CNRS), Conseil R\'{e}gional 
 Ile-de-France, D\'{e}partement Physique Nucl\'{e}aire 
 et Corpusculaire (PNC-IN2P3/CNRS), D\'{e}partement 
 Sciences de l'Univers (SDU-INSU/CNRS), Institut 
 Lagrange de Paris (ILP) Grant No. LABEX ANR-10-LABX-63, 
 within the Investissements d'Avenir Programme  
 Grant No. ANR-11-IDEX-0004-02, France; 
 Bundesministerium f\"{u}r Bildung und Forschung (BMBF), 
 Deutsche Forschungsgemeinschaft (DFG), 
 Finanzministerium Baden-W\"{u}rttemberg, 
 Helmholtz Alliance for Astroparticle Physics (HAP), 
 Helmholtz-Gemeinschaft Deutscher Forschungszentren (HGF), 
 Ministerium f\"{u}r Wissenschaft und Forschung, Nordrhein Westfalen, 
 Ministerium f\"{u}r Wissenschaft, Forschung und Kunst, Baden-W\"{u}rttemberg, Germany; 
 Istituto Nazionale di Fisica Nucleare (INFN), Ministero dell'Istruzione, dell'Universit\'{a} 
 e della Ricerca (MIUR), Gran Sasso Center for Astroparticle Physics (CFA), CETEMPS Center 
 of Excellence, Ministero degli Affari Esteri (MAE), Italy; 
 Consejo Nacional de Ciencia y Tecnolog\'{\i}a (CONACYT), Mexico; 
 Ministerie van Onderwijs, Cultuur en Wetenschap, 
 Nederlandse Organisatie voor Wetenschappelijk Onderzoek (NWO), 
 Stichting voor Fundamenteel Onderzoek der Materie (FOM), Netherlands; 
 National Centre for Research and Development, 
 Grants No. ERA-NET-ASPERA/01/11 and 
 No. ERA-NET-ASPERA/02/11, National Science Centre,
 Grants No. 2013/08/M/ST9/00322, No. 2013/08/M/ST9/00728 
 and No. HARMONIA 5 - 2013/10/M/ST9/00062, Poland; 
 Portuguese national funds and FEDER funds within 
 Programa Operacional Factores de Competitividade 
 through Funda\c{c}\~{a}o para a Ci\^{e}ncia e a  Tecnologia (COMPETE), Portugal; 
 Romanian Authority for Scientific Research ANCS, 
 CNDI-UEFISCDI partnership projects Grants No. 20/2012 
 and No. 194/2012, Grants No. 1/ASPERA2/2012 ERA-NET, 
 No. PN-II-RU-PD-2011-3-0145-17 and No. PN-II-RU-PD-2011-3-0062, 
 the Minister of National  Education, Programme  
 Space Technology and Advanced Research (STAR), 
 Grant No. 83/2013, Romania; 
 Slovenian Research Agency, Slovenia; 
 Comunidad de Madrid, FEDER funds, Ministerio de Educaci\'{o}n y Ciencia, 
 Xunta de Galicia, European Community 7th Framework Program, 
 Grant No. FP7-PEOPLE-2012-IEF-328826, Spain; 
 Science and Technology Facilities Council, United Kingdom; 
 Department of Energy, 
 Contracts No. DE-AC02-07CH11359, No. DE-FR02-04ER41300, 
 No. DE-FG02-99ER41107 and No. DE-SC0011689, 
 National Science Foundation, Grant No. 0450696, 
 The Grainger Foundation, USA; 
 NAFOSTED, Vietnam; 
 Marie Curie-IRSES/EPLANET, European Particle Physics 
 Latin American Network, European Union 7th Framework 
 Program, Grant No. PIRSES-2009-GA-246806; and UNESCO.

\section*{Pierre Auger Collaboration}
\begin{sloppypar}
A.~Aab$^{41}$,
P.~Abreu$^{67}$,
M.~Aglietta$^{52,53}$,
E.J.~Ahn$^{82}$,
I.~Al Samarai$^{28}$,
I.F.M.~Albuquerque$^{16}$,
I.~Allekotte$^{1}$,
P.~Allison$^{87}$,
A.~Almela$^{11,8}$,
J.~Alvarez Castillo$^{60}$
% ...
\end{sloppypar}

\vspace{1cm}
\par\noindent
$^{1}$ Centro At\'omico Bariloche and Instituto Balseiro (CNEA-UNCuyo-CONICET), San Carlos de Bariloche, Argentina\\
% ...
$^{8}$ Instituto de Tecnolog\'\i{}as en Detecci\'on y Astropart\'\i{}culas (CNEA, CONICET, UNSAM), Buenos Aires, Argentina\\
% ...
$^{11}$ Universidad Tecnol\'ogica Nacional -- Facultad Regional Buenos Aires, Buenos Aires, Argentina\\
% ...
$^{16}$ Universidade de S\~ao Paulo, Instituto de F\'\i{}sica, S\~ao Paulo, SP, Brazil\\
% ...
$^{28}$ Institut de Physique Nucl\'eaire d'Orsay (IPNO), Universit\'e Paris 11, CNRS-IN2P3, Orsay, France\\
% ...
$^{41}$ Universit\"at Siegen, Fachbereich 7 Physik -- Experimentelle Teilchenphysik, Siegen, Germany\\
% ...
$^{52}$ Osservatorio Astrofisico di Torino (INAF), Torino, Italy\\
$^{53}$ INFN, Sezione di Torino, Italy\\
% ...
$^{60}$ Universidad Nacional Aut\'onoma de M\'exico, M\'exico, D.F., M\'exico\\
% ...
$^{67}$ Laborat\'orio de Instrumenta\c{c}\~ao e F\'\i{}sica Experimental de Part\'\i{}culas (LIP) and Instituto Superior T\'ecnico, Universidade de Lisboa (UL), Portugal\\
% ...
$^{82}$ Fermilab, Batavia, IL, USA\\
% ...
$^{87}$ Ohio State University, Columbus, OH, USA\\
% ...


\begin{thebibliography}{30}
\bibitem {pao_nim} 
The Pierre Auger Collaboration, 
\emph{The Pierre Auger Cosmic Ray Observatory, } 
\emph{Nuclear Instruments and Methods in Physics Research, A \textbf{798} (2015) 172-213, }\href{http://arxiv.org/abs/1502.01323}{arXiv:1502.01323}

\bibitem{amiga_icrc} 
A. Etchegoyen for the Pierre Auger Collaboration.
\emph{AMIGA, Auger Muons and Infill for the Ground Array, }
\emph{30th ICRC, M\'erida, \textbf{5} (2007) 1191, }\href{http://arxiv.org/abs/0710.1646}{arXiv:0710.1646}

\bibitem{aera} 
Frank G. Schr\"oder for the Pierre Auger Collaboration
\emph{Radio detection of high-energy cosmic rays with the Auger Engineering Radio Array, }
\emph{Nuclear Instruments and Methods in Physics Research, A (2015), }\href{http://www.sciencedirect.com/science/article/pii/S0168900215009997}{doi:10.1016/j.nima.2015.08.047}

\bibitem{icrc:2013}
F. Suarez for the Pierre Auger Collaboration, 
\emph{The AMIGA muon detectors of the Pierre Auger Observatory: overview and status, }
\emph{33rd ICRC, Rio de Janeiro, (2013) \#712, }\href{http://arxiv.org/abs/1307.5059}{arXiv:1307.5059}

\bibitem{estudio_suelos} 
Report from Segermar Intemin, 
\emph{n6150 (27/03/2008)}, Av.Gral Paz entre Constituyentes y Nazca, Parque Tecnol\'ogico Miguelete - CC149. B1650 WAB, - San Mart\'in - Bs. As. Argentina (available by request).


\bibitem{paper_daniel} 
D. Supanitsky et al., 
\emph{Underground muon counters as a tool for composition analyses, }
\href{http://www.sciencedirect.com/science/article/pii/S0927650508000741}{\emph{Astroparticle Physics 29 (2008) 461-470.}}

\bibitem{icrc2011:modulos} 
M. Platino et al.,
\emph{Fabrication and testing system for plastic scintillator muon counters used in cosmic showers detection, }
\emph{32nd ICRC, Beijing (2011), \#0004.}
\href{http://www.ihep.ac.cn/english/conference/icrc2011/paper/proc/v4/v4_0008.pdf}{doi:10.7529/ICRC2011/V04/0004}

\bibitem{icrc2013:ale} 
A. Almela et al., 
\emph{Design and implementation of an embedded system for particle detectors, }
\href{http://www.cbpf.br/~icrc2013/papers/icrc2013-1209.pdf}{\emph{33rd ICRC, Rio de Janeiro (2013) \#1209.}}

\bibitem{scint_fermilab} 
A. Pla-Dalmau, A.D. Bross, V.V. Rykalin,
\emph{Extruding plastic scintillator at Fermilab, }
\emph{FERMILAB-Conf-03-318-E., }
\href{http://ieeexplore.ieee.org/xpl/abstractAuthors.jsp?arnumber=1352007}{doi:10.1109/NSSMIC.2003.1352007}

\bibitem{minos} 
D.G. Michael et al.,
\emph{The magnetized steel and scintillator calorimeters of the MINOS experiment, }
\emph{Nuclear Instruments and Methods in Physics Research, A. \textbf{596} (2008) 190-228, }
\href{http://arxiv.org/abs/0805.3170}{arXiv:0805.3170}

\bibitem{icrc2011:brian}
B. Wundheiler for the Pierre Auger Collaboration, 
\emph{The AMIGA muon counters of the Pierre Auger Observatory: performance and first data, }
\emph{32nd ICRC, Beijing (2011), \#0341. }
\href{http://arxiv.org/abs/1107.4807}{arXiv:1107.4807}

\bibitem{paper_diego} 
D. Ravignani and A.D. Supanitsky, 
\emph{A new method for reconstructing the muon lateral distribution with an array of segmented counters, }
\emph{Astroparticle Physics \textbf{65} (2015) 1-10. }
\href{http://arxiv.org/abs/1411.7649}{arXiv:1411.7649}

\bibitem{icrc_pmts} 
F. Suarez et al., 
\emph{A fully automated test facility for multi anode photo multiplier tubes, }
\emph{32nd ICRC, Beijing (2011) \#0020.}
\href{http://www.ihep.ac.cn/english/conference/icrc2011/paper/proc/v4/v4_0020.pdf}{doi:10.7529/ICRC2011/V04/0020}

\bibitem{agus_mux} 
A. Lucero et al., 
\emph{Analog multiplexer for testing multianode photomultipliers used in AMIGA project of the Pierre Auger Observatory, }
\emph{JINST (2015) T09004.}
\href{http://iopscience.iop.org/article/10.1088/1748-0221/10/09/T09004/meta;jsessionid=1E39556B1CA860CCC27DAA20DD950E13.c2.iopscience.cld.iop.org}{doi:10.1088/1748-0221/2015/9/T09004}

\bibitem{paper_scanner}
M. Platino et al., 
\emph{AMIGA at the Auger Observatory: the scintillator module testing system, }
\emph{JINST (2011) P06006. }
\href{http://iopscience.iop.org/1748-0221/6/06/P06006/}{doi:10.1088/1748-0221/6/06/P06006}
 
\bibitem{Opera}
T. Adam et al., 
\emph{The OPERA experiment Target Tracker, }
\emph{Nuclear Instruments and Methods in Physics Research, A. \textbf{577} (2007) 523-539. }
\href{http://arxiv.org/abs/physics/0701153}{	arXiv:physics/0701153}

\bibitem{Terlinsky:08} S. Terlisky, 
\emph{Unidad de Actividad Materiales, Gerencia de Investigaci\'on, Desarrollo y Asistencia T\'ecnica, CNEA (2008). } 
\href{http://www2.cnea.gov.ar/contacto/contactos.php}{http://www2.cnea.gov.ar/contacto/contactos.php}

\bibitem{jinst:oscar} 
O. Wainberg et al., 
\emph{Digital electronics for the Pierre Auger Observatory AMIGA muon counters, }
\emph{JINST \textbf{9} (2014) T04003. }
\href{http://arxiv.org/abs/1312.7131}{arXiv:1312.7131}

\bibitem{bubble_test} 
\emph{Standard test method for leaks using bubble emission techniques, }
\emph{ASTM International, West Conshohocken, PA, 2000. }
\href{http://www.astm.org/DATABASE.CART/HISTORICAL/E515-95R00.htm}{ASTM E515-95(2000).}

\end{thebibliography}
\end{document}